%% file: main.tex
\documentclass[final,5p,times,authoryear]{elsarticle}

\usepackage{microtype}

\usepackage{dblfloatfix} 
\usepackage{placeins}    
\usepackage{xcolor}

\usepackage{graphicx}
\usepackage{booktabs}
\usepackage{subcaption}
\usepackage{longtable}
\usepackage{array}
\usepackage{tabularx}
\usepackage{multirow}
\usepackage{makecell}
\usepackage{adjustbox}
\usepackage{threeparttable}
\usepackage{pdflscape}

\graphicspath{{figures/}{paper_figures/}{./}}

\usepackage{enumitem}
\setlist[itemize]{leftmargin=*, itemsep=2pt, topsep=2pt}
\setlist[enumerate]{leftmargin=*, itemsep=2pt, topsep=2pt}

\usepackage{amsmath}
\usepackage{amssymb}
\usepackage{siunitx}

\usepackage[T1]{fontenc}
\usepackage[utf8]{inputenc}

\usepackage{xurl}
\usepackage[hidelinks]{hyperref}

\newcommand{\bench}{AgriJetsonBench}
\newcommand{\agx}{Jetson AGX Orin}
\newcommand{\agxfull}{NVIDIA Jetson AGX Orin 64GB}
\newcommand{\nano}{Jetson Orin Nano Super}
\newcommand{\nanofull}{NVIDIA Jetson Orin Nano Super}

\newcommand{\yoloN}{YOLO11n}
\newcommand{\yoloS}{YOLO11s}
\newcommand{\rtdetr}{RT-DETR-R18}
\newcommand{\bisenet}{BiSeNetV2}
\newcommand{\mobilenet}{MobileNetV3-LRASPP}
\newcommand{\deeplab}{DeepLabV3+}
\newcommand{\segformer}{SegFormer-B0}

\newcommand{\jinf}{J/inference}
\newcommand{\whinf}{Wh/1000 inferences}
\newcommand{\infwh}{inferences/Wh}
\newcommand{\jgmac}{J/GMAC}

\newcommand{\matched}{matched-budget}
\newcommand{\nativeview}{native/high-performance}

\newif\ifwithsupplementary
\withsupplementaryfalse

\newif\ifdraftnotes
\draftnotesfalse

\journal{Artificial Intelligence in Agriculture}

\begin{document}

\begin{frontmatter}

\title{\bench: External-Power-Referenced TensorRT Benchmarking of Agricultural Vision Models on Jetson Edge Platforms}


\author[inst1]{Hasan Jahanifar\corref{cor1}}
\ead{hjahanifar@uga.edu}

\author[inst1]{Hasan Mirzakhaninafchi}
\ead{hasan.mirzakhaninafchi@uga.edu}

\author[inst2]{Wesley M. Porter}
\ead{wporter@uga.edu}

\author[inst3]{Abolfazl Najar}
\ead{najar@uga.edu}

\author[inst1,inst2]{Glen C. Rains}
\ead{grains@uga.edu}

\cortext[cor1]{Corresponding author}

\address[inst1]{College of Engineering, University of Georgia, Tifton, GA 31793, USA}
\address[inst2]{College of Agricultural and Environmental Sciences, University of Georgia, Tifton, GA 31793, USA}
\address[inst3]{College of Engineering, University of Georgia, Athens, GA 30602, USA}

\input{abstract}


\end{frontmatter}

\begin{figure*}[!t]
    \centering
    \IfFileExists{Fig1_workflow_AgriJetsonBench.pdf}
    {\includegraphics[width=0.98\textwidth]{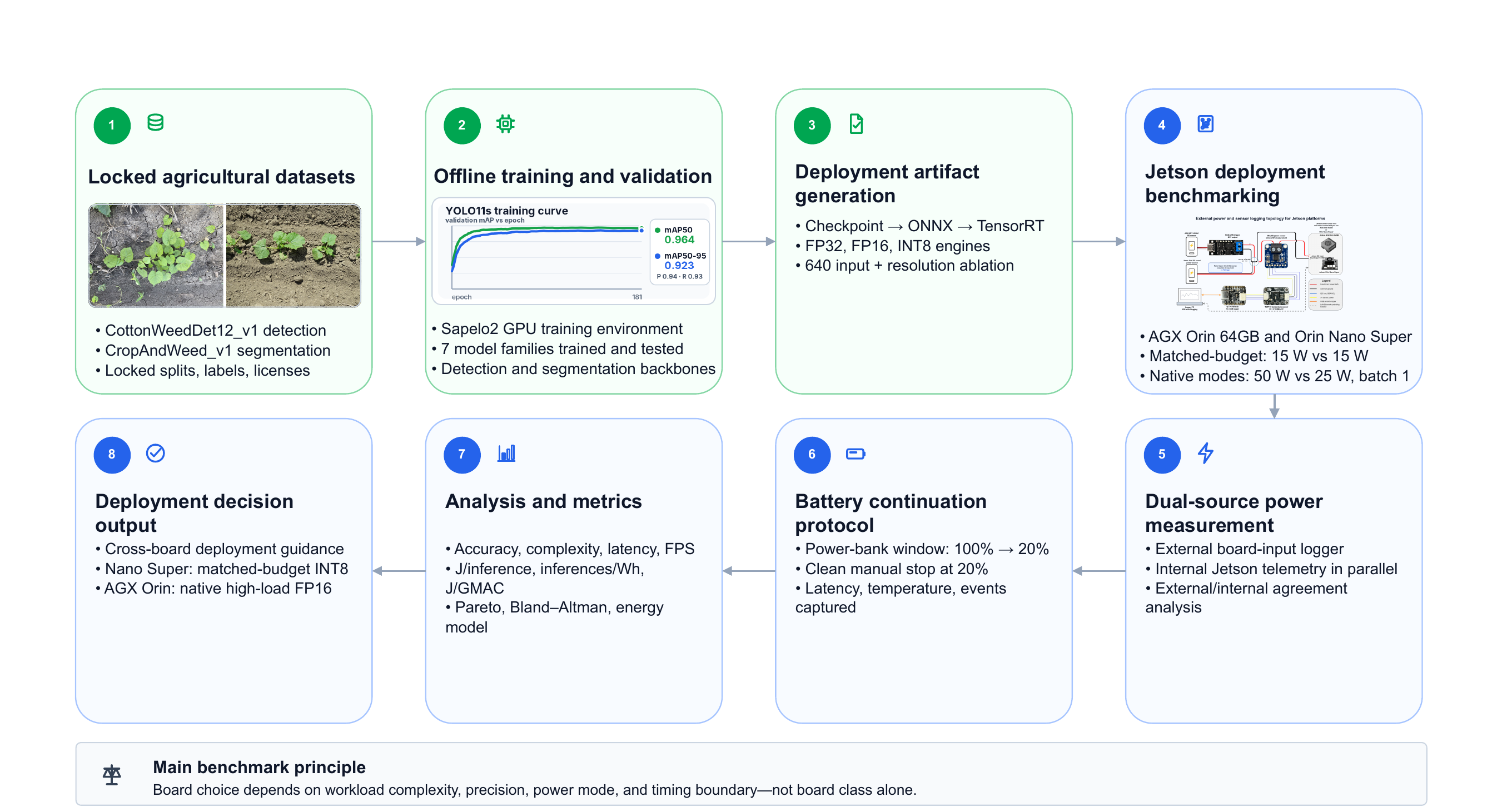}}
    {\fbox{\parbox{0.92\textwidth}{\centering Placeholder for Fig1\_workflow\_AgriJetsonBench.pdf.\\
    Workflow: dataset locking $\rightarrow$ training/validation $\rightarrow$ ONNX/TensorRT export $\rightarrow$ Jetson benchmarking $\rightarrow$ external/internal power analysis $\rightarrow$ deployment decision map.}}}
    \caption{AgriJetsonBench workflow. Green boxes indicate offline/pre-deployment stages, including dataset locking, model training, and deployment-artifact generation. Blue boxes indicate Jetson deployment, dual-source power measurement, battery continuation, analysis, and deployment-decision stages. Agricultural detection and segmentation datasets were locked and split, models were trained and validated offline, deployment artefacts were exported to ONNX and TensorRT, and Jetson inference was evaluated using external board-input power logging, internal telemetry, battery continuation tests, model-complexity analysis, and deployment decision mapping.}
    \label{fig:workflow}
\end{figure*}

\input{introduction}
\input{methods}
\input{results}
\input{discussion}
\input{conclusions}
\input{availability}
\input{declarations}


\FloatBarrier

\biboptions{authoryear,round}
\bibliographystyle{elsarticle-harv}
\bibliography{references}

\end{document}

%% file: abstract.tex

\begin{abstract}
Agricultural vision models are often selected from validation accuracy and reported frames per second, but deployment on embedded agricultural edge-GPU systems also depends on timing boundary, numeric precision, power mode, board-input energy, and sustained-run validity. We present AgriJetsonBench, a reproducible deployment benchmark for crop/weed detection and segmentation on NVIDIA Jetson AGX Orin 64GB and Jetson Orin Nano Super. Existing locked agricultural datasets were used as fixed deployment workloads, and seven model families were exported through ONNX and TensorRT. The benchmark combines pure TensorRT engine timing, external board-input power logging, internal Jetson telemetry, battery-continuation testing, run-validity screening, and archived reproducibility artifacts. Main energy tests used YOLO11s and SegFormer-B0 at batch size 1 and $640 \times 640$. Under matched 15~W INT8 operation, Orin Nano Super outperformed AGX Orin on both workloads: YOLO11s reached 145.82~FPS and 0.0767~J/inference versus 80.90~FPS and 0.1814~J/inference, and SegFormer-B0 reached 47.18~FPS and 0.2808~J/inference versus 31.95~FPS and 0.5013~J/inference. In native SegFormer-B0 FP16 operation, AGX Orin achieved higher throughput and lower p95 latency than Nano Super, 103.61~FPS and 9.441~ms versus 67.47~FPS and 15.020~ms, with similar energy per inference. Time-aligned analysis of 76,691 internal--external power samples showed high temporal association but a pooled internal-minus-external bias of $-1.988$~W. The results show that agricultural Jetson deployment decisions should be based on workload, precision, power mode, and measurement boundary rather than board class alone.
\end{abstract}

%% file: introduction.tex

\section{Introduction}
\label{sec:introduction}

Artificial intelligence is increasingly central to precision agriculture, where field decisions often depend on timely interpretation of visual data from cameras, robots, mobile platforms, and distributed sensing systems. Agricultural computer-vision models are now used for tasks such as weed detection, crop monitoring, disease assessment, livestock monitoring, yield estimation, and robotic perception \cite{liakos2018machine,kamilaris2018deep,shamshiri2018research}. In many of these applications, inference cannot rely exclusively on cloud computing because field connectivity may be limited, communication latency can be unacceptable, and raw image transmission may increase bandwidth, cost, and energy demand. Edge AI therefore has an important role in agricultural systems that require local autonomy, low latency, and energy-aware operation.

NVIDIA Jetson platforms are widely used for agricultural edge-AI prototyping because they provide GPU acceleration, embedded deployment support, and software compatibility with common deep-learning frameworks and TensorRT optimization tools \cite{nvidia_jetson_orin,nvidia_tensorrt}. However, selecting a Jetson platform and an agricultural vision model is not a simple accuracy-ranking problem. In field deployment, a model that performs well on a validation set may still be unsuitable if it has high latency, high tail latency, high memory pressure, excessive board-input power, or poor battery endurance. Conversely, a lightweight model may be attractive for a battery-powered system even if it is not the absolute accuracy leader. These trade-offs make agricultural edge-AI deployment a multi-objective decision problem.

Recent agricultural edge-AI studies show that embedded inference is practical for detection, segmentation, monitoring, and robotic perception tasks \cite{yoloweeds2023,aicropcam2023,rtal2023,tinysegformer2024,islam2025weed,wang2026livestock}. However, the reported deployment evidence remains difficult to compare across papers. FPS values often omit the timing boundary, power values may not specify whether they come from external board-input logging or internal telemetry, and platform comparisons rarely separate matched nominal-power operation from board-native high-performance operation. As a result, a validation-accuracy result cannot be reliably translated into a battery-aware Jetson deployment choice.

The overall \bench{} workflow is summarized in Fig.~\ref{fig:workflow}. In the workflow diagram, green boxes denote offline/pre-deployment stages and blue boxes denote Jetson deployment, measurement, analysis, and decision stages.

Table~\ref{tab:related_gap_mapping} summarizes this deployment-evidence gap for closely related agricultural edge-AI studies. The table is not intended as a ranking; it records whether each study reports the specific evidence needed for reproducible, energy-aware Jetson comparison.

\begin{table*}[t]
\centering
\caption{Gap mapping of closely related agricultural edge-AI studies.}
\label{tab:related_gap_mapping}
\begingroup
\scriptsize
\setlength{\tabcolsep}{2.0pt}
\renewcommand{\arraystretch}{1.12}
\begin{tabularx}{\textwidth}{@{}
>{\raggedright\arraybackslash}p{0.115\textwidth}
>{\raggedright\arraybackslash}p{0.105\textwidth}
>{\raggedright\arraybackslash}p{0.088\textwidth}
>{\raggedright\arraybackslash}p{0.108\textwidth}
>{\raggedright\arraybackslash}p{0.097\textwidth}
>{\raggedright\arraybackslash}p{0.087\textwidth}
>{\raggedright\arraybackslash}p{0.097\textwidth}
>{\raggedright\arraybackslash}p{0.087\textwidth}
>{\raggedright\arraybackslash}X
@{}}
\toprule
\textbf{Study} &
\textbf{Agricultural task} &
\textbf{Jetson / multi-board} &
\textbf{Numeric precision / quantization} &
\textbf{Power or energy} &
\textbf{Boundary stated} &
\textbf{Sustained / thermal} &
\textbf{Reproducibility assets} &
\textbf{Primary emphasis} \\
\midrule
\textit{YOLOWeeds}\newline\citep{yoloweeds2023} &
Cotton weed detection &
No Jetson comparison &
Not reported as deployment precision axis &
Not reported &
Model benchmark &
Not reported &
Code + dataset &
YOLO detector and dataset benchmark for multi-class cotton weed detection. \\
\addlinespace[1pt]
\textit{AICropCAM}\newline\citep{aicropcam2023} &
Crop monitoring: classification, segmentation, detection, counting &
No; Raspberry Pi / Arduino system &
Not reported as deployment precision axis &
System power reported; no J/inference &
Application pipeline &
Not thermal; field-oriented platform &
Partial &
Integrated edge camera system with speed and power reporting. \\
\addlinespace[1pt]
\textit{RTAL}\newline\citep{rtal2023} &
Rice lodging area assessment &
Single Jetson Xavier NX &
Not reported as deployment precision axis &
Not reported &
UAV area-throughput pipeline &
Partial sortie-scale runtime; no thermal/power trace &
Not reported &
Real-time UAV edge-computing method for large-area rice lodging mapping. \\
\addlinespace[1pt]
\textit{TinySegformer}\newline\citep{tinysegformer2024} &
Agricultural pest segmentation &
Single Jetson deployment context &
Partial; pruning / quantization, no FP32/FP16/INT8 energy protocol &
Not reported &
Model / device FPS &
Not reported &
Not reported &
Lightweight segmentation architecture with pruning/quantization for edge devices. \\
\addlinespace[1pt]
Islam et al.\newline\citep{islam2025weed} &
Weed detection and segmentation &
Partial; Jetson Nano and Orin Nano deployments &
Not reported as deployment precision axis &
Not reported &
Application-level edge FPS &
Not reported &
Data on request &
Real-time lightweight CNN workflow for weed detection and segmentation on edge devices. \\
\addlinespace[1pt]
Wang et al.\newline\citep{wang2026livestock} &
Livestock keypoint detection &
Partial; GPU / Jetson / CPU platforms &
Not reported as deployment precision axis &
Not reported &
Latency, FPS, memory, model size &
Not reported &
Code + dataset &
Multi-platform livestock pose benchmark with deployability metrics. \\
\addlinespace[1pt]
\textbf{This work} &
Crop/weed detection and segmentation &
\textbf{Yes; AGX Orin and Orin Nano Super} &
\textbf{FP32, FP16, and INT8 TensorRT where calibration succeeded} &
\textbf{External board-input J/inference and inferences/Wh} &
\textbf{Pure TensorRT engine-level and supplementary end-to-end evidence} &
\textbf{Battery continuation, temperature, events, idle baselines} &
\textbf{Tables, figures, scripts, manifests, checksums, logger traces, telemetry} &
\textbf{Unified matched-budget and native-mode Jetson benchmark for agricultural deployment decisions.} \\
\bottomrule
\multicolumn{9}{@{}>{\raggedright\arraybackslash}p{\textwidth}@{}}{\footnotesize\emph{Note.} ``Numeric precision / quantization'' refers to numeric deployment precision or quantization as an explicit benchmark axis, not the detection/segmentation precision metric. ``Partial'' means that the dimension is addressed in some form, but not as part of a unified, externally checked, dual-source, matched-budget agricultural Jetson benchmark protocol with explicit timing and board-input energy boundaries. ``Not reported'' means the dimension was not explicitly reported as a benchmark output in the cited study.}\\
\end{tabularx}
\endgroup
\end{table*}

This gap matters because edge-AI deployment decisions are affected by interactions among model architecture, input size, precision mode, batch size, power mode, and measurement boundary. A model may be efficient in INT8 at a matched 15 W budget but less favourable in FP16 native mode. A larger board may appear inefficient when underutilized in low-power batch-1 inference, but may become favourable when a heavier workload exposes its native compute capacity. Similarly, a reported energy value is difficult to compare unless the measurement boundary is explicit and the energy source is clearly defined. For this reason, agricultural edge-AI benchmarks need to report not only accuracy and FPS, but also external board-input power, energy per inference, compute-normalized energy, thermal behaviour, run validity, and reproducibility evidence.

Power measurement is a particularly important source of ambiguity. Jetson devices provide internal telemetry that is useful for diagnostics, temperature monitoring, and rail-level trends \cite{nvidia_tegrastats,nvidia_jetson_power}. However, internal telemetry does not necessarily equal total board-input energy, and rail coverage can differ by device and operating state. External inline power logging can provide a more direct board-input reference when it is inserted before the device power input \cite{ti_ina260}. For deployment studies, the most useful approach is to record both channels: external board-input logging as the primary energy reference and internal telemetry as a diagnostic and agreement-analysis channel. This dual-source design helps distinguish true board-input energy from internal rail estimates and enables transparent reporting of bias between the two. The same concern appears in broader embedded-ML benchmarking, where accuracy, latency, energy, hardware configuration, and measurement stack must be reported together for interpretable comparison \citep{banbury2021mlperf}. It is also consistent with Green AI arguments that predictive performance should be reported alongside computational cost rather than treated as the only optimization target \citep{schwartz2020green}.

Another source of ambiguity is the timing boundary. End-to-end image-pipeline measurements include image loading, decoding, preprocessing, postprocessing, non-maximum suppression, mask decoding, visualization, and storage overhead. These measurements are useful for complete application profiling, but they should not be mixed with pure inference-engine measurements. TensorRT pure-engine benchmarking isolates the optimized model execution path and supports controlled comparison across models, precisions, power modes, and Jetson platforms. Therefore, a reproducible benchmark should state whether it reports TensorRT engine-level inference, full image-pipeline inference, or both.

To address these needs, we developed \bench{}, a deployment-measurement benchmark for agricultural vision models on Jetson edge-AI platforms that links model complexity, TensorRT execution, and externally logged board-input energy. The benchmark connects offline model development to deployment measurement. It uses locked agricultural detection and segmentation datasets, trains and validates representative model families, exports deployment artifacts to ONNX and TensorRT, evaluates inference on two Jetson platforms, records external and internal power channels, performs power-bank continuation tests, and reports deployment decisions using latency, throughput, energy, model complexity, and run-validity metrics.

The benchmark evaluates two agricultural vision tasks: object detection and semantic segmentation. Detection models were trained on the locked \texttt{CottonWeedDet12\_v1} dataset, and segmentation models were trained on the locked \texttt{CropAndWeed\_v1} dataset \cite{cottonweeddet12_dataset,cropandweed_dataset}. The retained model set covered three detection models and four segmentation models; Section~\ref{sec:model_set} lists the individual architectures and their deployment roles. All accepted models were trained, validated, exported, and archived before Jetson deployment. Because this study uses existing agricultural image datasets, these datasets are treated as fixed deployment workloads for benchmarking rather than as newly collected agronomic datasets. The primary contribution is the externally referenced deployment-measurement protocol and measurement-boundary analysis for agricultural edge-computing systems.

The deployment study uses two Jetson boards: \agxfull{} and \nanofull{}. Two comparison views are separated throughout the paper. The matched-budget view compares AGX Orin 15 W against Nano Super 15 W. This view asks which board is preferable when both devices operate under the same nominal power class. The native/high-performance view compares AGX Orin 50 W against Nano Super 25 W. This view asks what each board can deliver in its practical high-performance mode. Keeping these views separate avoids overinterpreting either nominal-power fairness or board-native maximum performance.

The main latency and energy results are reported using a pure TensorRT engine-level boundary at batch size 1. This boundary was selected because the target use case is online agricultural inference, where frames arrive sequentially from cameras or field platforms and per-frame latency is deployment-relevant. Under this boundary, the reported energy metrics are J/inference, inferences/Wh, and Wh/1000 inferences. End-to-end image-pipeline evidence is retained only as supplementary context and is not mixed with the main TensorRT energy comparisons.

The main contributions of this paper are as follows:
\begin{enumerate}
    \item We define \bench{}, a training-to-deployment benchmark protocol for agricultural vision models that couples locked datasets, TensorRT deployment, external board-input power logging, internal telemetry, run-validity screening, and reproducibility packaging.
    \item We quantify accuracy, model complexity, artifact footprint, TensorRT latency, throughput, external energy per inference, inferences/Wh, and J/GMAC for representative detection and segmentation model families.
    \item We separate two deployment questions that are often conflated: matched-budget operation, where AGX Orin and Orin Nano Super are compared at 15 W, and native/high-performance operation, where each board is evaluated in its practical high-performance mode.
    \item We provide an artifact package containing split files, training metrics, model manifests, hardware/software metadata, power logs, telemetry, run-validity summaries, tables, figures, scripts, and SHA256 checksum manifests.
\end{enumerate}

The results show that the deployment winner is not fixed by board class. In matched-budget 15 W INT8 operation, Nano Super outperformed AGX Orin for both \yoloS{} and \segformer{}. In native high-load \segformer{} FP16 operation, AGX Orin delivered substantially higher throughput and lower latency while maintaining comparable energy per inference and compute-normalized energy. A native \rtdetr{} FP16 contrast case further shows that model architecture can change the energy ranking. Together, these findings support a practical deployment rule: board choice in agricultural edge AI should depend on workload complexity, precision, power mode, and measurement boundary, not on hardware class alone.

The remainder of this paper is organized as follows. Section~\ref{sec:methods} describes the datasets, model training, complexity profiling, Jetson hardware, TensorRT deployment, external and internal power measurement, battery continuation protocol, and energy metrics. Section~\ref{sec:results} presents training outcomes, complexity--accuracy analysis, locked-640 benchmark results, battery continuation results, internal--external power agreement, idle baselines, and deployment decision guidance. Section~\ref{sec:discussion} discusses the implications for agricultural edge-AI deployment, measurement validity, limitations, and future extensions. The final sections summarize the conclusions and describe data, code, and reproducibility availability.

%% file: methods.tex

\section{Materials and methods}
\label{sec:methods}

\subsection{Study overview and benchmark design}
\label{sec:methods_overview}

We developed \bench, a reproducible AI deployment benchmark for precision-agriculture vision models on NVIDIA Jetson edge platforms. The benchmark was designed to evaluate trained agricultural object-detection and semantic-segmentation models not only by predictive accuracy, but also by deployment-relevant behaviour: TensorRT latency, throughput, model complexity, board-input power, energy per inference, battery endurance, internal telemetry, thermal response, and run validity.

The experimental workflow followed the stages introduced in Fig.~\ref{fig:workflow}: dataset locking and split packaging, offline model training and validation, ONNX export and TensorRT engine generation, Jetson pure-engine benchmarking, and energy-aware deployment analysis.

The primary latency, throughput, and energy results use a pure TensorRT engine-level timing boundary. Unless explicitly stated otherwise, one ``inference'' refers to one batch-1 TensorRT engine invocation at the configured input size. The timed region excludes JPEG/image loading, dataset file I/O, CPU-side preprocessing, image resizing outside the engine, postprocessing, non-maximum suppression, mask decoding outside the engine, visualization, and prediction saving. Therefore, the main energy metrics are reported as \jinf, \infwh, and \whinf, rather than end-to-end J/image.

Two comparison views were used. The \matched{} view compares \agx{} in 15 W mode with \nano{} in 15 W mode. The \nativeview{} view compares \agx{} in 50 W mode with \nano{} in 25 W mode. This separation prevents fair nominal-power comparison from being conflated with the practical question of what each board provides in its board-native high-performance mode.

\subsection{Datasets and locked splits}
\label{sec:datasets}

Two agricultural vision tasks were evaluated: weed/crop object detection and crop/weed semantic segmentation. The detection task used the locked \texttt{CottonWeedDet12\_v1} dataset. The segmentation task used the locked \texttt{CropAndWeed\_v1} dataset.

The detection split contained 3,972 training images, 564 validation images, and 1,112 test images. The corresponding numbers of bounding boxes were 6,576, 908, and 1,886. The class list was frozen in the dataset package through \texttt{class\_names.txt}, \texttt{class\_names.json}, and the locked dataset YAML file.

The segmentation dataset contained three semantic classes:
\begin{equation*}
0:\ \text{background}, \qquad
1:\ \text{crop}, \qquad
2:\ \text{weed}.
\end{equation*}
The locked segmentation split contained 5,381 training images, 798 validation images, and 1,526 test images. The segmentation test package preserved paired image and mask lists, benchmark-order files, split metadata, class names, and dataset license notes.

\begin{table}[t]
\centering
\caption{Locked datasets and split sizes used in AgriJetsonBench.}
\label{tab:dataset_splits}
\scriptsize
\setlength{\tabcolsep}{3pt}
\begin{tabularx}{\columnwidth}{@{}>{\raggedright\arraybackslash}p{0.28\columnwidth}
>{\raggedright\arraybackslash}p{0.18\columnwidth}
r r r
>{\raggedright\arraybackslash}p{0.18\columnwidth}@{}}
\toprule
Dataset & Task & Train & Val & Test & Test annotations \\
\midrule
\texttt{CottonWeedDet12\_v1} & Detection & 3,972 & 564 & 1,112 & 1,886 boxes \\
\texttt{CropAndWeed\_v1} & Segmentation & 5,381 & 798 & 1,526 & 1,526 masks \\
\bottomrule
\end{tabularx}
\end{table}

Dataset split files, class names, split policies, license notes, and runtime manifests were stored in the reproducibility package under \texttt{raw\_evidence/dataset\_splits}. Raw images were not redistributed when dataset licenses restrict redistribution. The Jetson deployment tables can be recomputed from the archived engine metadata, run logs, latency files, and external power traces, whereas independent retraining requires access to the original datasets through their source providers.

\subsection{Offline training and validation workflow}
\label{sec:training_workflow}

Model training and validation were performed offline on the Sapelo2 GPU environment operated by the Georgia Advanced Computing Resource Center at the University of Georgia \citep{gacrc_sapelo2}. The Jetson boards were used for deployment benchmarking and energy measurement, not for training. The training workflow followed the same high-level structure for all model families:
\begin{enumerate}
    \item prepare locked dataset splits and runtime manifests;
    \item run a smoke-training check where applicable;
    \item perform hyperparameter tuning where applicable;
    \item train the full model;
    \item validate on the validation split;
    \item evaluate on the held-out test split;
    \item export the accepted model to ONNX;
    \item archive metrics, manifests, checksums, and selected qualitative outputs.
\end{enumerate}

The detection models were \yoloN{}, \yoloS{}, and \rtdetr{}. The YOLO models were trained using the Ultralytics training stack with Optuna tuning \citep{akiba2019optuna}. The RT-DETR model used a separate RT-DETR training and export workflow. The segmentation models were \bisenet{}, \mobilenet{}, \deeplab{}, and \segformer{}. Segmentation training used the locked three-class crop/weed dataset.

Training evidence was collected in a metrics-only archive and a figures-only archive. The metrics archive included final summaries, validation and test metrics, training arguments, tuning reports, ONNX export reports, output manifests, source snapshots, and SHA256 checksums. The figures archive included selected training curves, confusion matrices, precision--recall curves, per-class plots, and validation batch visualizations. Heavy model weights and raw datasets were not required for manuscript-level tables, but model files were tracked by path, size, and SHA256 hash.

\subsection{Model set}
\label{sec:model_set}

The benchmark retained seven trained model families spanning lightweight detection, larger detection, lightweight segmentation, conventional CNN segmentation, and transformer-style segmentation:
\begin{itemize}
    \item \yoloN{} and \yoloS{} for YOLO-family object detection;
    \item \rtdetr{} for non-YOLO detection;
    \item \bisenet{} and \mobilenet{} for lightweight semantic segmentation;
    \item \deeplab{} for heavier CNN-based segmentation;
    \item \segformer{} for transformer-style semantic segmentation.
\end{itemize}

The main cross-board battery analysis focuses on \yoloS{} and \segformer{}. \yoloS{} represents a practical lightweight detector, while \segformer{} represents a higher-load segmentation workload with strong predictive accuracy. \rtdetr{} is retained as a native-mode contrast case because it shows that board ranking can change with model architecture. \bisenet{} is included in the input-resolution ablation as a lightweight segmentation representative.

\subsection{Model complexity profiling}
\label{sec:model_complexity}

Model complexity was reported alongside accuracy and deployment metrics. For each retained model, we recorded parameter count, MACs at 640 $\times$ 640, GMACs, GFLOPs-equivalent values under the convention $1$ MAC $=2$ FLOPs, checkpoint size, and ONNX artifact size. The complexity reference input was:
\[
\text{batch size}=1, \qquad \text{input tensor}=1\times 3\times 640\times 640.
\]

Table~\ref{tab:model_complexity} lists the complexity values used for all complexity-normalized metrics. These values were used to compute GMAC/s and J/GMAC:
\begin{equation}
\mathrm{GMAC/s} = C_m \times \mathrm{FPS},
\end{equation}
\begin{equation}
\mathrm{J/GMAC} = \frac{\mathrm{J/inference}}{C_m},
\end{equation}
where $C_m$ is the model complexity in GMACs per 640 inference.

\begin{table*}[t]
\centering
\caption{Model complexity and artifact footprint at 640 $\times$ 640.}
\label{tab:model_complexity}
\scriptsize
\setlength{\tabcolsep}{4pt}
\begin{adjustbox}{max width=\textwidth}
\begin{tabular}{l l r r r r l}
\toprule
Model & Task & Parameters & MACs at 640 & GMACs & GFLOPs equiv. & Artifact footprint \\
\midrule
\yoloN{} & Detection & 2.63 M & $3.76\times 10^{9}$ & 3.76 & 7.53 & checkpoint 5.2 MiB; ONNX 10.1 MiB \\
\yoloS{} & Detection & 9.46 M & $1.18\times 10^{10}$ & 11.81 & 23.61 & checkpoint 18.3 MiB; ONNX 36.2 MiB \\
\rtdetr{} & Detection & 20.02 M & $3.08\times 10^{10}$ & 30.82 & 61.63 & checkpoint 307.5 MiB; ONNX 76.6 MiB \\
\bisenet{} & Segmentation & 3.33 M & $1.92\times 10^{10}$ & 19.23 & 38.47 & checkpoint 39.9 MiB; ONNX 12.8 MiB \\
\mobilenet{} & Segmentation & 3.21 M & $3.25\times 10^{9}$ & 3.25 & 6.50 & checkpoint 24.8 MiB; ONNX 12.3 MiB \\
\deeplab{} & Segmentation & 40.32 M & $1.08\times 10^{11}$ & 108.37 & 216.74 & checkpoint 326.2 MiB; ONNX 153.8 MiB \\
\segformer{} & Segmentation & 3.71 M & $1.48\times 10^{10}$ & 14.84 & 29.67 & checkpoint 42.7 MiB; ONNX 14.3 MiB \\
\bottomrule
\end{tabular}
\end{adjustbox}
\end{table*}

Model complexity was used for interpretation and normalized metrics, not as a substitute for measured deployment performance. TensorRT kernel fusion, precision mode, memory bandwidth, clock policy, and board power management can change the relationship between theoretical MACs and measured latency or energy.

\subsection{Jetson hardware platforms}
\label{sec:hardware}

Two NVIDIA Jetson development kits were used:
\begin{itemize}
    \item \agxfull{} Developer Kit;
    \item \nanofull{} Developer Kit.
\end{itemize}

The AGX Orin was treated as the higher-performance edge platform, and the Orin Nano Super as the lower-power edge platform. Two deployment views were defined as follows:
\begin{center}
\footnotesize
\begin{tabular}{@{}lcc@{}}
\toprule
View & AGX Orin & Nano Super \\
\midrule
matched-budget & 15 W & 15 W \\
native/high-performance & 50 W & 25 W \\
\bottomrule
\end{tabular}
\end{center}

Power-mode IDs and mode names were confirmed before experiments using:
\begin{quote}
\footnotesize\ttfamily
sudo nvpmodel -q\\
sudo nvpmodel -q --verbose
\end{quote}
The exact software stack, JetPack/L4T version, CUDA version, TensorRT version, Python package versions, active nvpmodel mode, and clock policy were recorded in hardware/software metadata tables. Optional Nano MAXN\_SUPER evidence was not merged into the main native comparison; if included, it is treated only as supplementary ceiling evidence.

\subsection{Model export and TensorRT deployment}
\label{sec:tensorrt_export}

Each accepted trained model was exported through the following deployment path:
\begin{center}
\footnotesize
\begin{tabular}{c}
trained checkpoint \\
$\downarrow$ \\
ONNX export \\
$\downarrow$ \\
TensorRT engine \\
$\downarrow$ \\
Jetson benchmark
\end{tabular}
\end{center}

For each model, input size, precision, and batch-size configuration, a separate TensorRT engine was generated. Static engines were not reused across input resolutions. The evaluated precision modes were FP32, FP16, and TensorRT INT8 build mode. FP32 served as a reference precision where supported, and FP16 represented the main optimized deployment reference. INT8 calibration was input-size-specific; a 640 calibration cache was not reused for other resolutions. For the main 640 cross-board engines used in the battery-energy comparison, each INT8 engine used 256 calibration images recorded in the corresponding local calibration manifest. Calibration-cache paths, TensorRT build commands/logs, engine metadata, engine SHA256 checksums, evaluator commands, and metric JSON files were archived with the corresponding run evidence.

INT8 accuracy retention was checked for the main 640 cross-board engines, namely YOLO11s and SegFormer-B0 on AGX Orin and Orin Nano Super, against the corresponding FP16 TensorRT engine on the same board, input size, and batch-size configuration. For YOLO11s detection, INT8 was accepted when the absolute drops in COCO mAP50--95, mAP50, and AR100 were each no greater than 0.03 and no important class AP collapse was observed. The TensorRT COCO evaluator did not emit a standalone scalar precision value, so scalar precision was not used as a retention criterion. For SegFormer-B0 segmentation, INT8 was accepted when the absolute drops in mIoU and Dice were no greater than 0.05, crop and weed IoU drops were no greater than 0.10, and no foreground class collapsed. SegFormer-B0 INT8 engines allowed TensorRT FP16 fallback where required by the builder. This retention check is scoped to the main 640 cross-board INT8 engines and is not claimed as an exhaustive retention evaluation for every supplementary INT8 engine or resolution-ablation output.

\subsection{Primary timing boundary}
\label{sec:timing_boundary}

The primary benchmark boundary was pure TensorRT engine-level inference. The timed region excluded:
\begin{itemize}
    \item JPEG/image loading and dataset file I/O;
    \item CPU-side preprocessing and normalization;
    \item postprocessing and non-maximum suppression;
    \item mask decoding outside the engine;
    \item visualization and prediction saving;
    \item framework-specific per-image path overhead.
\end{itemize}

Pure-engine timing used either \texttt{trtexec\_random\_input} or a preloaded TensorRT engine runner. Both were accepted as pure-engine measurements when the timed region contained only batch-1 TensorRT inference. When random-input timing was used, it was interpreted only as TensorRT engine-execution timing and energy; predictive accuracy was taken from the locked validation and test evaluations, not from random-input runs. End-to-end image-pipeline runs were archived as supplementary evidence, but they were not mixed with the main pure-engine tables because they measure a different boundary.

\subsection{Locked 640 benchmark matrix}
\label{sec:locked_640}

The main benchmark matrix was locked at an input size of 640 $\times$ 640 and batch size 1. Batch size 1 was selected because the target deployment scenario is online agricultural inference, such as frame-by-frame processing from a camera mounted on a field robot, mobile inspection platform, or edge sensor node. This setting preserves directly interpretable per-frame median and p95 latency.

The locked 640 matrix included all seven model families and up to three precision modes. AGX locked-640 results included matched 15 W and native 50 W modes. Nano Super locked-640 and battery evidence was merged into the cross-board paper tables where available. Each row recorded model, task, board, power mode, precision, input size, timing boundary, latency, FPS, external power, energy, memory, temperature, and run-validity fields.

\subsection{Input-resolution ablation}
\label{sec:resolution_ablation}

Input-resolution experiments were performed as a supporting ablation, not as a replacement for the locked 640 matrix. The ablation identified:
\begin{itemize}
    \item $S_{\min}$: the smallest accuracy-preserving input size;
    \item $S_{\max}$: the largest memory-safe common native input size across both Jetson boards.
\end{itemize}

For detection, candidate sizes were:
\[
320, 416, 512, 640, 768, 896, 1024, 1280, 1536.
\]
For segmentation, candidate sizes were:
\[
256, 320, 384, 448, 512, 576, 640, 704, 768, 896, 1024, 1152.
\]

The 640 FP16 result was used as the reference. A smaller detection size was accepted only if the absolute drops in mAP50-95, mAP50, and recall were each no greater than 0.03, and no important class suffered a large AP collapse. A smaller segmentation size was accepted only if mIoU and Dice dropped by no more than 0.05 absolute, crop and weed IoU did not drop by more than 0.10 absolute, and no foreground class collapsed.

Maximum-size discovery used FP16 and the native/high-performance mode of each board: AGX 50 W and Nano Super 25 W. A large size was retained as $S_{\max}$ only if it was memory-safe on both boards under these native modes. A large size passed only if the TensorRT engine built successfully, a short smoke run passed, no CUDA OOM occurred, no Linux OOM-killer event was observed, swap use remained zero or negligible, and peak RAM stayed below the safe memory threshold. Selected sizes were then confirmed with matched 15 W smoke tests.

\begin{sloppypar}
The full official resolution-power ablation was restricted to four representative models: \yoloS{}, \rtdetr{}, \bisenet{}, and \segformer{}.
\end{sloppypar}

\subsection{External power measurement topology}
\label{sec:external_power}

A central methodological feature of \bench{} is dual-source power measurement. The external inline logger measured board-input voltage, current, and power before the Jetson input and was treated as the primary energy reference. Internal Jetson telemetry was recorded in parallel but used only as a diagnostic and agreement-analysis channel.

The upstream power-source stage was board- and run-type-specific. For adapter-powered runs, the original board-compatible power adapters or sources were used upstream of the external logger. The AGX Orin adapter/source path was routed through a 20 V USB-C PD trigger before entering the external logger. The Orin Nano Super adapter/direct-input path was routed upstream of the same external logger without a PD trigger, because the Nano Super input path did not require a separate PD trigger under that setup. For power-bank continuation tests, the upstream source was the UGREEN 25,000 mAh 145 W USB-C PD power bank; because the power-bank output was USB-C PD, a 20 V PD trigger was used upstream of the external logger whenever needed to provide the board-compatible DC input, including Nano Super power-bank runs. In all cases, the selected board-input source was routed through the external INA260 logger immediately before the Jetson board input.

The external logging assembly used an INA260 inline power sensor for board-input voltage, current, and power measurement and a TMP119 sensor for local temperature monitoring. Both sensors were read by an Adafruit QT Py RP2040 over I2C and streamed to the logger PC over USB serial. The logger PC also maintained a LAN/Ethernet connection to the Jetson for run control and log transfer; this LAN link was not used for energy integration.

The INA260 external logger was checked using a Chroma 62150H-600 programmable DC power supply and a Chroma 63804 programmable AC/DC electronic load. The test matrix included no-load voltage checks at 5, 12, 20, and 30 V, followed by constant-current load points from approximately 0.28 to 4.03 A near the Jetson input-voltage range. Each step was logged for 30 s through the same INA260--QT Py RP2040 USB-serial path used for benchmark logging. Across step means, the maximum absolute voltage and current differences were 57.9 mV and 7.40 mA, respectively; across loaded steps, the maximum absolute power difference was 0.354 W, corresponding to 0.533\% of the reference power at the worst step. Linear voltage and current correction models were archived as characterization evidence, while the manuscript energy tables use the external logger traces consistently across all benchmark configurations.

Thus, the measurement hierarchy was:
\[
\text{external inline logger} = \text{primary board-input energy reference},
\]
\[
\text{internal Jetson telemetry} = \text{diagnostic and agreement-analysis channel}.
\]

\begin{figure*}[t]
    \centering
    \IfFileExists{Fig2_power_measurement_topology.pdf}
    {\includegraphics[width=\textwidth]{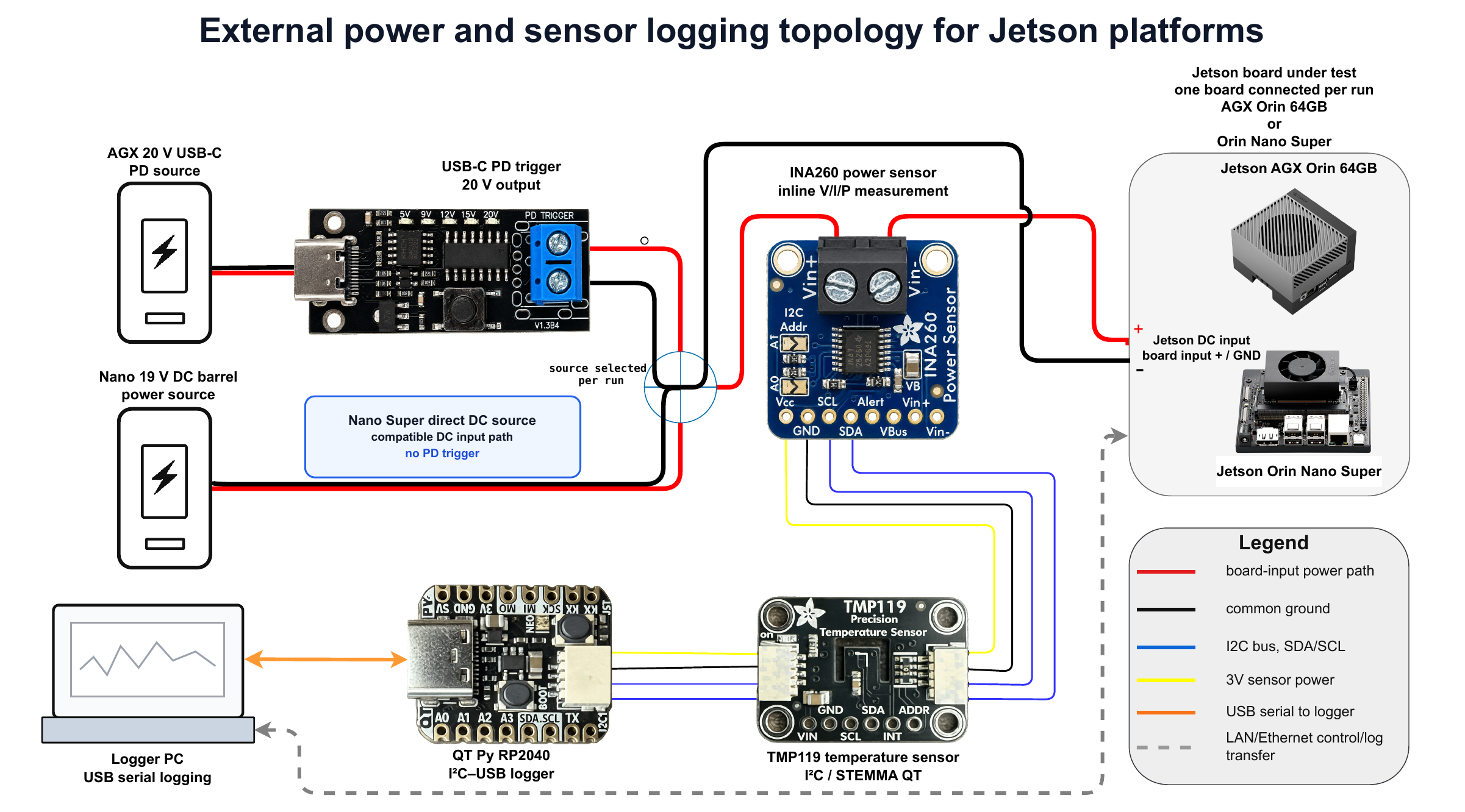}}
    {\fbox{\parbox{0.92\textwidth}{\centering Placeholder for Fig2\_power\_measurement\_topology.pdf.\\
    External board-input logging with board-specific upstream source path, INA260 inline sensor, QT Py RP2040 logger bridge, internal telemetry, and logger PC.}}}
    \caption{External power and sensor logging topology for the Jetson AGX Orin and Jetson Orin Nano Super platforms. The upstream power-source stage differed by board and run type: AGX Orin adapter/source runs used a 20 V USB-C PD trigger before the external logger, Nano Super adapter/direct-input runs used a compatible direct input path without a PD trigger, and USB-C power-bank continuation runs used a 20 V PD trigger where required by the power-bank source. In all cases, the INA260 inline sensor measured board-input voltage, current, and power immediately before the Jetson board input. The Adafruit QT Py RP2040 read the INA260 and TMP119 sensors over I2C and streamed records to the logger PC over USB serial. The dashed LAN/Ethernet link was used only for Jetson run control and log transfer, not for energy integration.}
    \label{fig:power_topology}
\end{figure*}

The external logger recorded raw voltage, current, and power samples at approximately 10 Hz. This raw external logger stream was the source for board-input energy integration. For internal--external agreement analysis only, external and internal power were converted to a separate paired 1 Hz overlap table; this table is not the raw external sampling rate. The power-bank display was used only to define the battery stop point; it was not used to compute energy. Figure~\ref{fig:power_topology} summarizes the measurement topology.

\subsection{Internal telemetry}
\label{sec:internal_telemetry}

Internal telemetry was collected using Jetson-native monitoring outputs, including tegrastats-derived power, temperature, memory, process, and rail-level information where available. The internal telemetry stream was used to estimate maximum internal temperature, memory pressure, and internal power trends, and to quantify agreement with the external logger.

Internal telemetry was not treated as the primary energy source because rail coverage and aggregation differ across boards and operating modes. Instead, the measurement hierarchy was:
\[
\text{external logger} = \text{primary board-input energy reference},
\]
\[
\text{internal telemetry} = \text{diagnostic and agreement-analysis channel}.
\]

\subsection{Battery continuation protocol}
\label{sec:battery_protocol}

Long-duration battery continuation tests were used to evaluate sustained deployment behaviour under a field-like energy source. Each main battery run followed the same procedure:
\begin{enumerate}
    \item charge the UGREEN power bank to 100\%;
    \item connect the board-specific power-source path through the external logger;
    \item boot the Jetson from the power bank;
    \item confirm the active nvpmodel mode;
    \item confirm that the external logger reports the expected board-input voltage and nonzero power;
    \item start internal telemetry logging;
    \item enter the noise-guard screening context where supported;
    \item start the pure-engine TensorRT benchmark;
    \item continue until the power-bank display reaches 20\% remaining;
    \item stop the run cleanly using manual Ctrl-C;
    \item restore the noise-guard context after the run where applicable;
    \item save summaries, latency files, event records, noise-guard summaries, sanitized run-validity summaries, external logger archives, internal telemetry, and stop annotations.
\end{enumerate}

Because the UGREEN power bank output was USB-C PD, a 20 V PD trigger was used upstream of the external logger when needed to provide the board-compatible DC input for the power-bank continuation runs, including the Nano Super power-bank runs. The PD trigger was part of the power-bank input path and not part of the energy computation; all energy values were computed from the external logger.

All main battery continuation tests covered the displayed 100\% to 20\% power-bank state-of-charge window and were stopped at 20\% remaining. This display window was used only as a standardized stop condition; runs were not allowed to continue to power-bank cutoff or Jetson reboot. The standardized stop reason was:
\begin{quote}
\footnotesize\ttfamily
manual\_ctrl\_c\_power\_bank\_80\_percent\_used\_\\
20\_percent\_remaining
\end{quote}

The main continuation configurations were:
\begin{itemize}
    \item \yoloS{} INT8, 640, AGX 15 W and Nano Super 15 W;
    \item \segformer{} INT8, 640, AGX 15 W and Nano Super 15 W;
    \item \segformer{} FP16, 640, AGX 50 W and Nano Super 25 W.
\end{itemize}

\subsection{Noise-guard screening and run validity criteria}
\label{sec:validity}

Deployment runs used a best-effort noise-guard screening context when supported by the run scripts. The guard was designed to reduce common non-benchmark background activity while preserving the external logger/dashboard route. In the safe profile, it could disable Wi-Fi when the logger route did not require it, block Bluetooth, blank the local display, stop selected background services such as package update and discovery services, and then restore the prior state after the run. The guard wrote JSON summaries for enter and restore actions; these summaries were treated as diagnostic evidence and were not used to compute latency or energy. Noise-guard screening was used together with the validity checks below to exclude or flag runs affected by benchmark failures or unstable measurement conditions.

A benchmark run was considered valid only if all required configuration and measurement conditions were satisfied:
\begin{itemize}
    \item the intended board and nvpmodel mode were confirmed;
    \item the intended model, precision, input size, and batch size were used;
    \item the timing boundary matched the planned pure-engine or supplementary boundary;
    \item the external logger was active and external energy was integrable;
    \item external voltage was recorded throughout each run and checked for consistency with the board-compatible upstream source path described in Section~\ref{sec:external_power}; no main battery-continuation run was excluded for a voltage-validity failure;
    \item logger gaps greater than 1 s were counted and reported; the six main battery-continuation runs had zero such gaps;    
    \item noise-guard summaries did not report fatal guard errors where the guard was used;
    \item run-validity summaries and diagnostic status records did not indicate benchmark failure;
    \item failed inference count was zero;
    \item no unexpected reboot, OOM event, or overtemperature shutdown occurred;
    \item battery tests used the standardized 100\% to 20\% stop rule;
    \item alarm events were saved and reported transparently.
\end{itemize}

Non-fatal current or high-power alarms did not automatically invalidate a run if the benchmark completed cleanly, external energy was valid, no overtemperature event occurred, and no benchmark failure was present. Such runs were reported as valid with diagnostic warnings. Among the six main battery-continuation runs, the Nano Super native 25 W SegFormer-B0 FP16 run was the one reported with current/high-power diagnostic warnings. For that run, the archived diagnostic thresholds were 1.25 A for overcurrent warning, 1.50 A for overcurrent critical, 25.0 W for high-power warning, and 30.0 W for high-power critical. These flags were used to annotate the run and did not by themselves invalidate it.

\subsection{Measured energy integration}
\label{sec:energy_integration}

External board-input energy was calculated from the external logger power trace. Let $P_{\mathrm{ext}}(t)$ denote external board-input power in watts and $N$ the number of timed TensorRT inferences. The external logger energy window, \(E_{\mathrm{Wh}}\), expressed in watt-hours, was computed as:
\begin{equation}
E_{\mathrm{Wh}} = \frac{1}{3600}\int_{t_0}^{t_1}P_{\mathrm{ext}}(t)\,dt .
\label{eq:energy_integral}
\end{equation}

For discrete logger samples:
\begin{equation}
E_{\mathrm{Wh}} \approx \frac{1}{3600}\sum_{i=1}^{K}P_i\Delta t_i .
\label{eq:energy_discrete}
\end{equation}

The main energy-normalized metrics were:
\begin{equation}
\mathrm{J/inference} = \frac{3600E_{\mathrm{Wh}}}{N},
\label{eq:j_per_inf}
\end{equation}
\begin{equation}
\mathrm{inferences/Wh} = \frac{N}{E_{\mathrm{Wh}}},
\label{eq:inf_per_wh}
\end{equation}
\begin{equation}
\mathrm{Wh}/1000~\mathrm{inferences} =
\frac{1000E_{\mathrm{Wh}}}{N}.
\label{eq:wh_per_1000}
\end{equation}

Throughput was calculated as:
\begin{equation}
\mathrm{FPS} = \frac{N}{t_1-t_0}.
\label{eq:fps}
\end{equation}

When both full-logger and inference-aligned energy windows were available, the inference-aligned external energy window was used for the main normalized metrics. Full-logger energy was retained only for sensitivity and traceability.

\subsection{Internal--external power agreement analysis}
\label{sec:agreement_methods}

For agreement analysis only, the external logger trace and internal telemetry stream were aligned over overlapping timestamp intervals using the archived paper-analysis workflow in \path{scripts/build_agreement_figures_and_decision_map.py}. External logger power and internal telemetry power were resampled to a common 1 Hz UTC timeline using mean power within each 1 s bin. The final agreement table was created by joining the external and internal 1 Hz series on matching timestamps within the overlapping run intervals, and rows were retained only for bins where both external and internal power values were available. This produced the archived \path{paper_tables/internal_external_time_aligned_samples_1hz.csv} table with 76,691 paired 1 Hz overlap samples across the six main battery-continuation runs. The table was used to characterize internal--external bias and temporal association; the main energy metrics remained based on the raw external logger power trace and the corresponding logger or inference-aligned energy window described in Section~\ref{sec:energy_integration}.

For each aligned sample $i$, the difference was:
\begin{equation}
d_i = P_{\mathrm{internal},i} - P_{\mathrm{external},i}.
\end{equation}

Bias, mean absolute error, and root mean square error were computed as:
\begin{equation}
\mathrm{bias} = \frac{1}{n}\sum_{i=1}^{n}d_i,
\end{equation}
\begin{equation}
\mathrm{MAE} = \frac{1}{n}\sum_{i=1}^{n}|d_i|,
\end{equation}
\begin{equation}
\mathrm{RMSE} =
\sqrt{\frac{1}{n}\sum_{i=1}^{n}d_i^2}.
\end{equation}

Pearson and Spearman correlations were also computed. Bland--Altman limits of agreement were calculated as:
\begin{equation}
\mathrm{LoA} = \overline{d} \pm 1.96s_d,
\end{equation}
where $\overline{d}$ is the mean difference and $s_d$ is the standard deviation of the differences. The agreement analysis was used to characterize internal telemetry bias and variability; it did not replace external board-input energy in the main results.

\subsection{Idle baselines}
\label{sec:idle_baselines}

Four short idle baseline runs were collected to estimate board/mode baseline external power $P_0$:
\begin{itemize}
    \item AGX 15 W idle;
    \item AGX 50 W idle;
    \item Nano Super 15 W idle;
    \item Nano Super 25 W idle.
\end{itemize}

\begin{sloppypar}
Each idle run used the same external logger and board-specific upstream power-source logic described in Section~\ref{sec:external_power}. Adapter-powered idle runs used the board-compatible adapter/source path upstream of the external logger; power-bank-specific PD triggering was used only when the upstream source required it. After setting the target nvpmodel mode and allowing a settling period, a 5-minute idle window was recorded with external logger and internal telemetry active. No TensorRT inference, Python benchmark, file copy, package installation, or model workload was allowed during the idle window.
\end{sloppypar}

The external average power during the idle window was used as $P_{0,b,r}$ for board $b$ and mode $r$ in the empirical energy model.

\subsection{Complexity-aware empirical energy model}
\label{sec:energy_model}

A complexity-aware empirical energy model was used to interpret deployment behaviour beyond raw FPS and \jinf{}. For model $m$ at input size $S$, complexity was represented as:
\begin{equation}
C_m(S) = C_{m,640}\left(\frac{S}{640}\right)^{q_m},
\label{eq:complexity_scaling}
\end{equation}
where $C_m(S)$ is GMACs per inference, $C_{m,640}$ is the measured GMACs at 640, and $q_m$ is an empirical input-scaling exponent. For the main locked 640 analysis, $C_m(S)=C_{m,640}$.

The compute rate was:
\begin{equation}
R = C_m(S)\times \mathrm{FPS},
\label{eq:compute_rate}
\end{equation}
with units GMAC/s. External power was modeled as:
\begin{equation}
\hat{P}_{\mathrm{ext}} =
P_{0,b,r} + \eta_{b,r,p}R,
\label{eq:power_model}
\end{equation}
where $P_{0,b,r}$ is the idle baseline power for board $b$ and mode $r$, and $\eta_{b,r,p}$ is an empirical dynamic energy slope for board $b$, mode $r$, and precision $p$.

The predicted energy per inference was:
\begin{equation}
\hat{J}_{\mathrm{inf}}
=
\frac{\hat{P}_{\mathrm{ext}}}{\mathrm{FPS}}
=
\frac{P_{0,b,r}}{\mathrm{FPS}}+\eta_{b,r,p}C_m(S).
\label{eq:jinf_model}
\end{equation}

This expression separates amortized platform overhead from workload-dependent compute energy. The term $P_0/\mathrm{FPS}$ becomes large when a board is underutilized or constrained by a low-power mode, while $\eta C_m$ represents the workload-dependent energy component. The model was used as an interpretive deployment estimator; all primary energy values were measured directly from the external logger.

\subsection{Statistical and reporting conventions}
\label{sec:statistical_reporting}

For short repeated benchmark runs, latency was summarized using median and p95 latency, and throughput and power were summarized using average values across repeats where available. For long battery continuation runs, each run represented a sustained deployment trial from 100\% to 20\% power-bank state of charge and was reported descriptively using total inferences, elapsed time, FPS, latency, external energy, average and peak external power, \jinf{}, \infwh{}, \whinf{}, maximum internal temperature, alarm status, and validity status.

The final interpretation used Pareto-style and deployment-decision views rather than a single accuracy-only leaderboard. Configurations were compared by accuracy, FPS, latency, \jinf{}, \jgmac{}, thermal stability, and power-mode context.

\subsection{Reproducibility package}
\label{sec:reproducibility_package}

All evidence was archived in \texttt{AgriJetsonBench\_v1}. The package includes:
\begin{itemize}
    \item dataset split files, class names, and license notes;
    \item training summaries, validation/test metrics, training reproducibility summaries, and artifact manifests;
    \item model complexity summaries at 640;
    \item Jetson hardware/software version outputs;
    \item TensorRT engine metadata and model artifact manifests;
    \item battery run summaries, latency CSV files, event records, noise-guard summaries, sanitized run-validity summaries, and external logger archives;
    \item internal telemetry logs and time-aligned internal--external agreement tables;
    \item idle baseline summaries;
    \item energy-model validation tables;
    \item paper figures, supplementary tables, scripts, and manifest files.
\end{itemize}

Each run was assigned a unique run ID and a paper-use category indicating whether it belonged to the main matched-budget results, the main native-mode results, a supplementary contrast case, or a supplementary non-comparable boundary case. Package-level manifests and SHA256 checksums were used to make the benchmark auditable and to support independent re-analysis of the reported tables.

%% file: results.tex

\section{Results}
\label{sec:results}

\subsection{Training, validation, and export outcomes}
\label{sec:results_training}

All seven target model families were successfully trained, validated, exported, and archived before Jetson deployment. The detection models were trained on \texttt{CottonWeedDet12\_v1}, and the segmentation models were trained on \texttt{CropAndWeed\_v1}. Table~\ref{tab:training_accuracy_results} summarizes the main validation and test metrics.

For object detection, \yoloN{} and \yoloS{} achieved similar test accuracy, with mAP50-95 values of 0.9035 and 0.9019, respectively. \rtdetr{} achieved a lower but still competitive test mAP50-95 of 0.8740, with COCO-style test AR@100 of 0.961. For semantic segmentation, \segformer{} achieved the strongest predictive performance, with test mIoU of 0.8049, test Dice of 0.8854, and foreground mIoU of 0.7115. \deeplab{} followed with test mIoU of 0.7774, while \mobilenet{} and \bisenet{} achieved 0.7631 and 0.7278, respectively.

\begin{table*}[t]
\centering
\caption{Training and held-out test accuracy for the retained models. Detection models are reported using AP/mAP metrics; YOLO rows also include scalar precision and recall from the retained evaluation summaries, while RT-DETR-R18 includes COCO-style AR@100 where available. Segmentation models are reported using mIoU, Dice, foreground mIoU, crop IoU, and weed IoU.}
\label{tab:training_accuracy_results}
\scriptsize
\setlength{\tabcolsep}{3pt}
\begin{adjustbox}{max width=\textwidth}
\begin{tabular}{l l l r r r r r r r r r r r}
\toprule
Model & Task & Dataset & Val primary & Test primary & Test mAP50 & Test AR100 & Precision & Recall & Test Dice & FG mIoU & Crop IoU & Weed IoU & GMACs \\
\midrule
\yoloN{} & Detection & CottonWeedDet12 & 0.9298 & 0.9035 & 0.9573 & -- & 0.9474 & 0.9091 & -- & -- & -- & -- & 3.76 \\
\yoloS{} & Detection & CottonWeedDet12 & 0.9303 & 0.9019 & 0.9591 & -- & 0.9593 & 0.9069 & -- & -- & -- & -- & 11.81 \\
\rtdetr{} & Detection & CottonWeedDet12 & 0.9180 & 0.8740 & 0.9290 & 0.961 & -- & -- & -- & -- & -- & -- & 30.82 \\
\bisenet{} & Segmentation & CropAndWeed & 0.6785 & 0.7278 & -- & -- & -- & -- & 0.8291 & 0.5972 & 0.6486 & 0.5457 & 19.23 \\
\mobilenet{} & Segmentation & CropAndWeed & 0.7431 & 0.7631 & -- & -- & -- & -- & 0.8560 & 0.6498 & 0.7009 & 0.5986 & 3.25 \\
\deeplab{} & Segmentation & CropAndWeed & 0.7393 & 0.7774 & -- & -- & -- & -- & 0.8664 & 0.6705 & 0.7179 & 0.6230 & 108.37 \\
\segformer{} & Segmentation & CropAndWeed & 0.8177 & 0.8049 & -- & -- & -- & -- & 0.8854 & 0.7115 & 0.7665 & 0.6565 & 14.84 \\
\bottomrule
\end{tabular}
\end{adjustbox}
\vspace{2pt}
\begin{minipage}{0.98\textwidth}
\footnotesize
Note: For detection models, ``primary'' denotes AP/mAP50-95. RT-DETR-R18 was evaluated using COCO-style AP/AR metrics; its reported Test AR100 is AR at IoU=0.50:0.95, area=all, maxDets=100. YOLO precision and recall are scalar values from the retained YOLO evaluation summaries and are not directly interchangeable with COCO AR@100.
\end{minipage}
\end{table*}

All retained models had traceable training and export evidence. For each model, the paper package includes a final summary, validation/test metrics, an exported ONNX model, checkpoint and ONNX SHA256 values, output manifests, and model artifact manifests. YOLO11n and YOLO11s each included six artifact manifest files; RT-DETR-R18 and the segmentation models included output manifests and checksum evidence. The ONNX export status was recorded as exported for all seven models.

\begin{table*}[t]
\centering
\caption{Training reproducibility and artifact traceability summary. Manuscript tables use summarized metrics and artifact metadata rather than embedding full binary model files; checkpoints and ONNX exports were tracked by path, size, and SHA256 hash.}
\label{tab:training_reproducibility}
\scriptsize
\setlength{\tabcolsep}{5pt}
\begin{adjustbox}{max width=\textwidth}
\begin{tabular}{l l r r r r r}
\toprule
Model & Task & Checkpoint (MiB) & ONNX (MiB) & Artifact manifests & SHA256 content lines & Export status \\
\midrule
\yoloN{} & Detection & 5.24 & 10.12 & 6 & 1139 & exported \\
\yoloS{} & Detection & 18.31 & 36.19 & 6 & 512 & exported \\
\rtdetr{} & Detection & 307.52 & 76.59 & 2 & 717 & exported \\
\bisenet{} & Segmentation & 39.86 & 12.75 & 2 & 2608 & exported \\
\mobilenet{} & Segmentation & 24.79 & 12.28 & 2 & 792 & exported \\
\deeplab{} & Segmentation & 326.22 & 153.84 & 2 & 1075 & exported \\
\segformer{} & Segmentation & 42.70 & 14.29 & 2 & 1021 & exported \\
\bottomrule
\end{tabular}
\end{adjustbox}
\end{table*}

\subsection{Model complexity versus predictive accuracy}
\label{sec:results_complexity_accuracy}

\begin{sloppypar}
The retained models covered a wide complexity range. \mobilenet{} and \yoloN{} were the smallest models by GMACs, at 3.25 and 3.76 GMACs per 640 inference, respectively. \deeplab{} was the largest model, with 108.37 GMACs and a 153.84 MiB ONNX artifact. \rtdetr{} was the largest detection model by checkpoint size and required 30.82 GMACs per 640 inference. \segformer{} had only 3.71 M parameters, but required 14.84 GMACs, making it a moderate-parameter but computationally meaningful segmentation workload.
\end{sloppypar}

\begin{figure}[t]
    \centering
    \IfFileExists{Fig3_model_complexity_vs_accuracy.pdf}
    {\includegraphics[width=\linewidth]{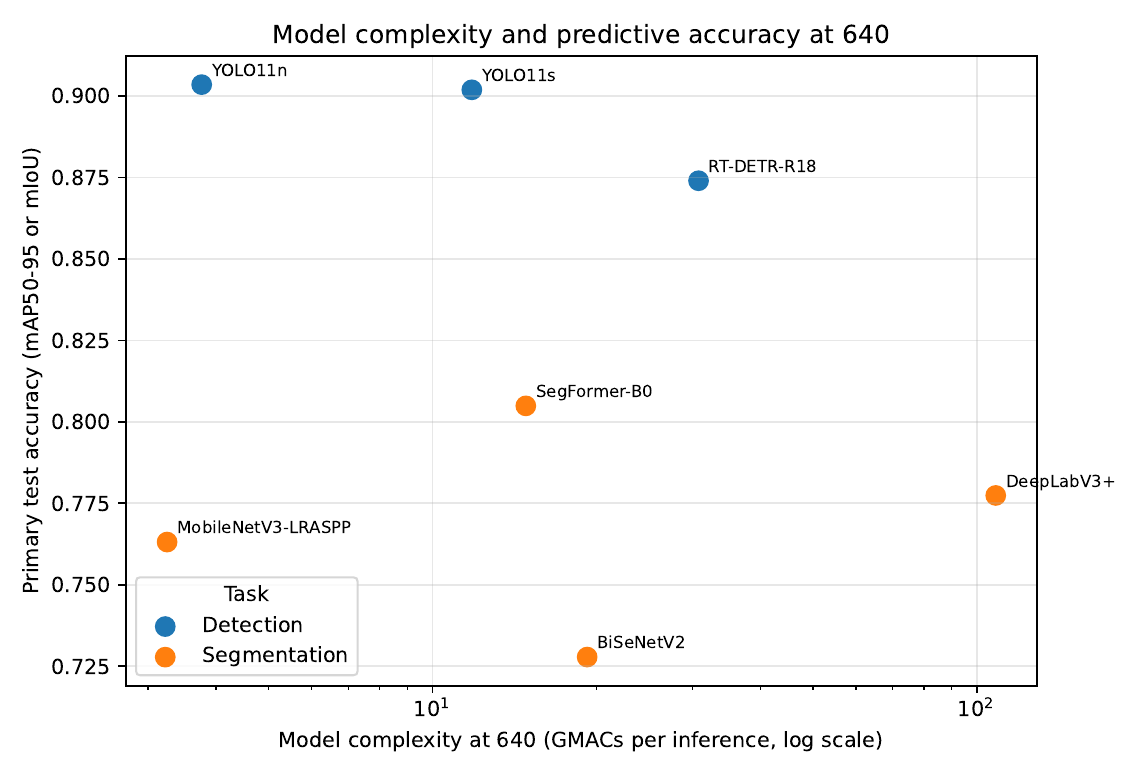}}
    {\IfFileExists{Fig3_model_complexity_vs_accuracy.png}
    {\includegraphics[width=\linewidth]{Fig3_model_complexity_vs_accuracy.png}}
    {\fbox{\parbox{0.92\linewidth}{\centering Placeholder for Fig3\_model\_complexity\_vs\_accuracy.}}}}
    \caption{Model complexity versus primary test accuracy. Detection models are evaluated using mAP50-95 and segmentation models using mIoU. The plot shows that parameter count and GMACs alone do not determine predictive performance, motivating joint accuracy--complexity--energy analysis.}
    \label{fig:complexity_accuracy}
\end{figure}

Figure~\ref{fig:complexity_accuracy} compares model complexity with primary test accuracy. The result shows that the most computationally expensive model was not necessarily the most accurate. \deeplab{} required much higher computation than \segformer{} but did not exceed it in segmentation mIoU. Similarly, \rtdetr{} required more GMACs than \yoloS{} but had lower test mAP50-95. This motivated the use of deployment-aware metrics rather than accuracy-only model ranking.

\subsection{Locked 640 benchmark matrix on AGX Orin}
\label{sec:results_agx_640}

The locked 640 AGX benchmark matrix contained 45 rows across seven model folders: \deeplab{}, \mobilenet{}, \bisenet{}, \rtdetr{}, \segformer{}, \yoloN{}, and \yoloS{}. The matrix included FP16, FP32, and INT8 precisions where supported, with matched 15 W and native 50 W AGX power tags.

Table~\ref{tab:agx_best_640_results} summarizes the best AGX locked-640 pure-engine result per model group. INT8 was the fastest and most energy-efficient precision for all seven AGX model groups. \yoloN{} INT8 achieved the lowest energy per inference, 0.0431 J/inference, while \deeplab{} INT8 remained the most expensive of the listed INT8 configurations at 0.2369 J/inference. These results show that precision optimization can substantially affect deployment efficiency, but the best deployment choice still depends on task, accuracy, model complexity, and board operating mode.

\begin{table*}[t]
\centering
\caption{Best AGX Orin locked-640 pure-engine result per model group. The original AGX table used an engine-level energy boundary; values are reported here as J/inference.}
\label{tab:agx_best_640_results}
\small
\setlength{\tabcolsep}{7pt}
\begin{tabular}{l r r r}
\toprule
Model group & Best precision & Median latency (ms) & J/inference \\
\midrule
\deeplab{} & INT8 & 8.353 & 0.2369 \\
\mobilenet{} & INT8 & 3.975 & 0.0800 \\
\bisenet{} & INT8 & 7.573 & 0.1269 \\
\rtdetr{} & INT8 & 6.499 & 0.1635 \\
\segformer{} & INT8 & 7.029 & 0.2083 \\
\yoloN{} & INT8 & 2.434 & 0.0431 \\
\yoloS{} & INT8 & 3.390 & 0.0685 \\
\bottomrule
\end{tabular}
\end{table*}

\subsection{Input-resolution ablation}
\label{sec:results_resolution}

The input-resolution ablation remained separate from the locked 640 benchmark matrix. Table~\ref{tab:resolution_selected_sizes} summarizes the selected input sizes for the four representative ablation models. \yoloS{} accepted a smaller accuracy-preserving size of 416 and scaled up to 1536. \bisenet{} accepted a smaller size of 448 and scaled up to 1152. \rtdetr{} and \segformer{} remained at 640 under the predefined accuracy and memory-safety rules.

\begin{table}[t]
\centering
\caption{Selected cross-board input sizes from the resolution ablation. $S_{\min}$ is the smallest accuracy-preserving input size relative to the 640 FP16 reference; $S_{\max}$ is the largest memory-safe common native FP16 input size across AGX Orin 50 W and Orin Nano Super 25 W. The main benchmark matrix remains locked at 640.}
\label{tab:resolution_selected_sizes}
\footnotesize
\setlength{\tabcolsep}{6pt}
\begin{tabular}{l l r r r}
\toprule
Model & Task & Reference & $S_{\min}$ & $S_{\max}$ common \\
\midrule
YOLO11s & Detection & 640 & 416 & 1536 \\
RT-DETR-R18 & Detection & 640 & 640 & 640 \\
BiSeNetV2 & Segmentation & 640 & 448 & 1152 \\
SegFormer-B0 & Segmentation & 640 & 640 & 640 \\
\bottomrule
\end{tabular}
\end{table}

For the selected AGX native 50 W power runs, the best energy-throughput combinations were obtained with INT8 at the accepted sizes. \yoloS{} INT8 at 416 achieved 518.33 FPS and 0.0435 J/inference. \bisenet{} INT8 at 448 achieved 212.05 FPS and 0.0870 J/inference. \rtdetr{} INT8 at 640 achieved 164.61 FPS and 0.1735 J/inference. \segformer{} INT8 at 640 achieved 156.19 FPS and 0.2110 J/inference. Therefore, smaller input sizes improved deployment efficiency for some models, but the benefit was model-dependent rather than universal.

\subsection{Battery continuation run validity}
\label{sec:results_validity}

All six main battery continuation runs were valid under the final paper criteria. Each used a pure-engine timing boundary, had external logger coverage, had zero logger gaps greater than 1 s, had available external energy, had no failed inferences, no reboot, no OOM, and no overtemperature event. All were reported using the final standardized stop reason:
\begin{quote}
\footnotesize\ttfamily
manual\_ctrl\_c\_power\_bank\_80\_percent\_used\_\\
20\_percent\_remaining
\end{quote}

The Nano Super native 25 W \segformer{} FP16 run was the only main battery-continuation run reported with current/high-power diagnostic warnings. It recorded 7663 external overcurrent-warning hits at the 1.25 A warning threshold and 8013 high-power-warning hits at the 25.0 W warning threshold. However, the archived run-validation summary indicated no error-channel failure, no failed inferences, no overtemperature event, and no tegrastats fault text. It was therefore retained as a valid native-mode run with power/current diagnostic warnings rather than as a failed run.

All energy formula checks passed. For each battery row, FPS, J/inference, inferences/Wh, and Wh/1000 inferences matched the reported inference count, elapsed time, and external energy window.

\subsection{Main battery continuation results}
\label{sec:results_battery_overview}

Table~\ref{tab:battery_main_results} summarizes the six main pure-engine battery continuation runs. All runs used batch size 1 and the 100\% to 20\% power-bank continuation protocol.

\begin{table*}[t]
\centering
\caption{Main pure-engine battery continuation results. All runs used batch size 1 and were stopped cleanly at 20\% power-bank remaining. Energy is based on external board-input logging.}
\label{tab:battery_main_results}
\scriptsize
\setlength{\tabcolsep}{3pt}
\begin{adjustbox}{max width=\textwidth}
\begin{tabular}{l l l l r r r r r r r r}
\toprule
Board & Mode & Model & Prec. & Inferences & Runtime (h) & FPS & p95 (ms) & Avg W & Energy (Wh) & J/inf. & Inf./Wh \\
\midrule
AGX Orin & 15 W & \yoloS{} & INT8 & 1,136,477 & 3.902 & 80.90 & 11.895 & 14.54 & 57.254 & 0.1814 & 19,850 \\
Nano Super & 15 W & \yoloS{} & INT8 & 2,760,205 & 5.258 & 145.82 & 7.018 & 11.12 & 58.843 & 0.0767 & 46,908 \\
AGX Orin & 15 W & \segformer{} & INT8 & 414,130 & 3.601 & 31.95 & 29.566 & 16.01 & 57.663 & 0.5013 & 7,182 \\
Nano Super & 15 W & \segformer{} & INT8 & 756,122 & 4.452 & 47.18 & 21.582 & 13.15 & 58.974 & 0.2808 & 12,821 \\
AGX Orin & 50 W & \segformer{} & FP16 & 660,564 & 1.771 & 103.61 & 9.441 & 33.20 & 58.807 & 0.3205 & 11,233 \\
Nano Super & 25 W & \segformer{} & FP16 & 683,695 & 2.815 & 67.47 & 15.020 & 21.79 & 61.331 & 0.3229 & 11,148 \\
\bottomrule
\end{tabular}
\end{adjustbox}
\end{table*}

The two matched-budget INT8 comparisons favoured Nano Super. The native/high-performance \segformer{} FP16 comparison favoured AGX Orin in throughput and latency, while energy per inference was nearly tied.

\subsection{Matched-budget 15 W INT8 result: YOLO11s}
\label{sec:results_yolo_matched}

In the matched 15 W comparison, Nano Super clearly outperformed AGX Orin for \yoloS{} INT8. Nano Super processed 2,760,205 inferences over 5.258 h, whereas AGX processed 1,136,477 inferences over 3.902 h. Nano achieved 145.82 FPS compared with 80.90 FPS on AGX, a 1.80$\times$ throughput advantage. Median latency decreased from 11.828 ms on AGX to 6.990 ms on Nano, and p95 latency decreased from 11.895 ms to 7.018 ms.

Energy efficiency also favoured Nano. AGX required 0.1814 J/inference, while Nano required 0.0767 J/inference. This corresponds to a 57.7\% reduction in energy per inference and a 2.36$\times$ increase in inferences/Wh. Because the two runs used comparable external energy windows, the Nano advantage reflects higher throughput and lower external power rather than only a longer runtime.

Using the \yoloS{} complexity value of 11.81 GMACs per inference, AGX achieved 955.1 GMAC/s and 0.01536 J/GMAC. Nano achieved 1721.7 GMAC/s and 0.00650 J/GMAC. Thus, Nano was not only lower power; it delivered more \yoloS{} computation per second and required less compute-normalized energy.

\subsection{Matched-budget 15 W INT8 result: SegFormer-B0}
\label{sec:results_segformer_matched}

The matched-budget \segformer{} INT8 comparison also favoured Nano Super. AGX processed 414,130 inferences over 3.601 h at 31.95 FPS, whereas Nano processed 756,122 inferences over 4.452 h at 47.18 FPS. Nano therefore achieved a 1.48$\times$ throughput advantage. Median latency decreased from 29.523 ms on AGX to 21.514 ms on Nano, and p95 latency decreased from 29.566 ms to 21.582 ms.

Energy efficiency again favoured Nano. AGX required 0.5013 J/inference, while Nano required 0.2808 J/inference, a 44.0\% reduction. Inferences/Wh increased from 7,182 on AGX to 12,821 on Nano.

The compute-normalized result was consistent with the raw energy result. Using the \segformer{} complexity value of 14.84 GMACs per inference, AGX achieved 474.0 GMAC/s and 0.03378 J/GMAC, whereas Nano achieved 700.0 GMAC/s and 0.01892 J/GMAC.

The AGX run was stable rather than failed. Median and p95 latency were tightly grouped, and maximum internal temperature was 52.06$^\circ$C. The result therefore indicates underutilization of the larger AGX platform in the 15 W batch-1 INT8 regime, not a run validity problem.

INT8 retention checks were performed for the four main 640 cross-board INT8 engines used in the battery-energy comparison. All four main INT8 engines passed the aggregate retention criteria relative to their board-matched FP16 TensorRT references. The check was scoped to YOLO11s and SegFormer-B0 at 640 on AGX Orin and Orin Nano Super, and was not intended as an exhaustive retention statement for every supplementary INT8 engine or resolution-ablation output.

\subsection{Native/high-performance result: SegFormer-B0 FP16}
\label{sec:results_segformer_native}

In the native/high-performance comparison, AGX Orin 50 W and Nano Super 25 W were evaluated using \segformer{} FP16 at 640. This workload better exposed the AGX platform's native throughput capacity than the matched-budget INT8 tests.

AGX achieved 103.61 FPS compared with 67.47 FPS on Nano, giving AGX a 1.54$\times$ throughput advantage. Median latency decreased from 14.842 ms on Nano to 9.420 ms on AGX, and p95 latency decreased from 15.020 ms to 9.441 ms. AGX therefore clearly won throughput and latency in the native \segformer{} FP16 regime.

The normalized energy result was more nuanced. Using inference-aligned external energy, AGX required 0.3205 J/inference, while Nano required 0.3229 J/inference. The two boards were therefore nearly tied in energy per inference, with only a marginal AGX advantage. Compute-normalized energy showed the same pattern: AGX achieved 1537.2 GMAC/s and 0.02160 J/GMAC, whereas Nano achieved 1001.0 GMAC/s and 0.02177 J/GMAC.

Thus, the strongest conclusion from the native \segformer{} FP16 battery test is that AGX delivered substantially higher throughput and lower latency without a meaningful energy-per-inference or J/GMAC penalty.

\subsection{Native RT-DETR-R18 FP16 contrast case}
\label{sec:results_rtdetr_contrast}

\rtdetr{} FP16 provided a useful contrast to the \segformer{} native result. In native mode, AGX was faster, reaching approximately 105.1 FPS with p95 latency of approximately 9.73 ms, while Nano reached approximately 72.7 FPS with p95 latency of approximately 13.79 ms. However, Nano remained more energy-efficient for this model, requiring approximately 0.270 J/inference compared with approximately 0.311 J/inference on AGX.

This contrast case shows that model complexity alone does not determine energy efficiency. \rtdetr{} has higher GMACs than \segformer{}, yet its native-mode energy ranking differed. Architecture, TensorRT execution behaviour, memory access patterns, precision mode, and board power policy all influence the final deployment trade-off.

\subsection{Pareto view of throughput and energy}
\label{sec:results_pareto}

\begin{figure}[t]
    \centering
    \IfFileExists{Fig4_fps_vs_j_inference_pareto.pdf}
    {\includegraphics[width=\linewidth]{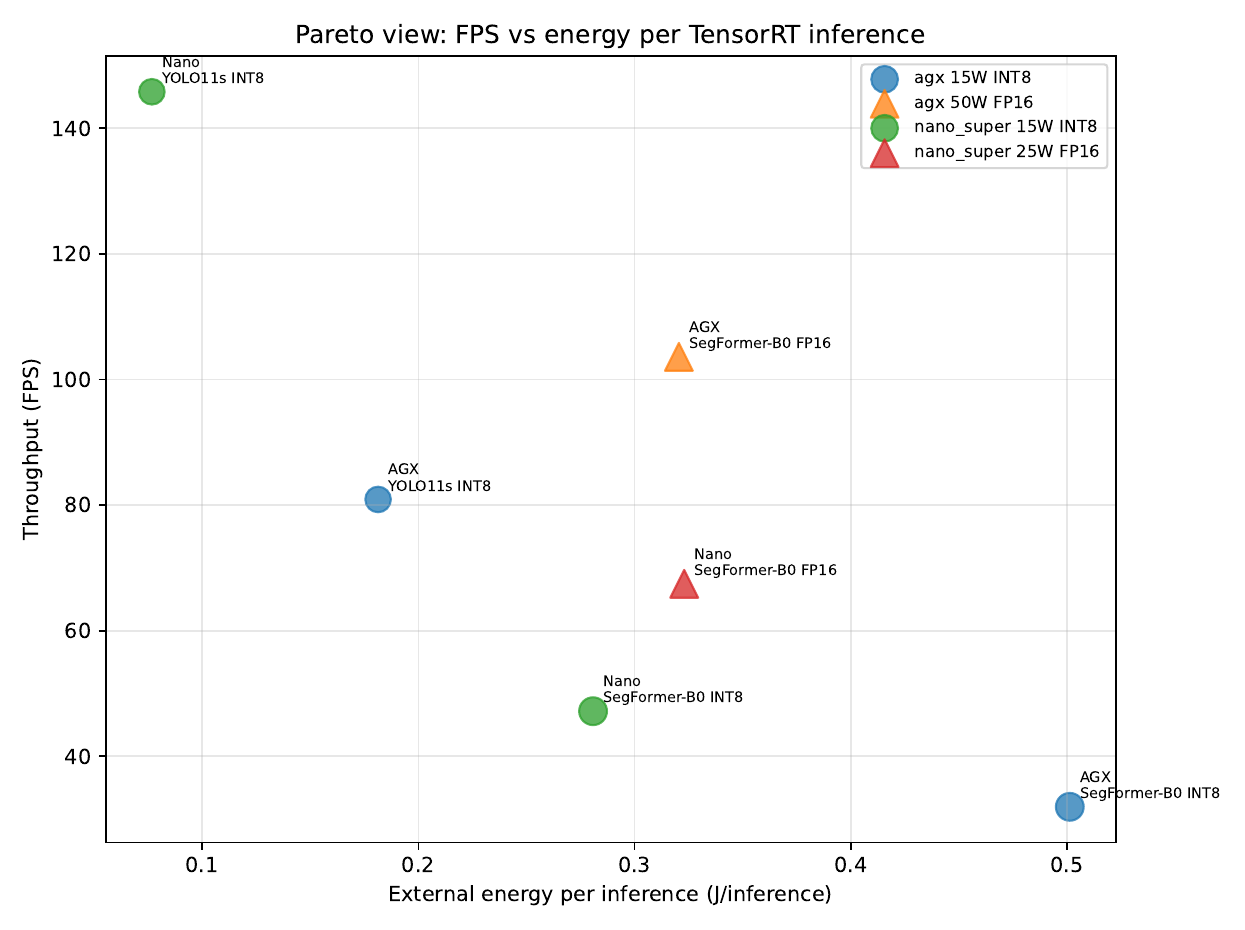}}
    {\IfFileExists{Fig4_fps_vs_j_inference_pareto.png}
    {\includegraphics[width=\linewidth]{Fig4_fps_vs_j_inference_pareto.png}}
    {\fbox{\parbox{0.92\linewidth}{\centering Placeholder for Fig4\_fps\_vs\_j\_inference\_pareto.}}}}
    \caption{Pareto view of FPS versus J/inference for the main battery configurations. Marker size represents model complexity in GMACs. Nano Super dominates the matched-budget INT8 configurations, while AGX provides the native high-load \segformer{} FP16 throughput advantage.}
    \label{fig:fps_jinf}
\end{figure}

Figure~\ref{fig:fps_jinf} summarizes the main battery configurations in the FPS versus J/inference plane. The matched-budget Nano Super INT8 points occupy the favourable high-FPS, low-energy region for \yoloS{} and \segformer{}. AGX moves to the high-throughput region only under native high-load \segformer{} FP16.

\subsection{Compute-normalized energy view}
\label{sec:results_jgmac}

\begin{figure}[t]
    \centering
    \IfFileExists{Fig5_j_gmac_vs_gmac_s.pdf}
    {\includegraphics[width=\linewidth]{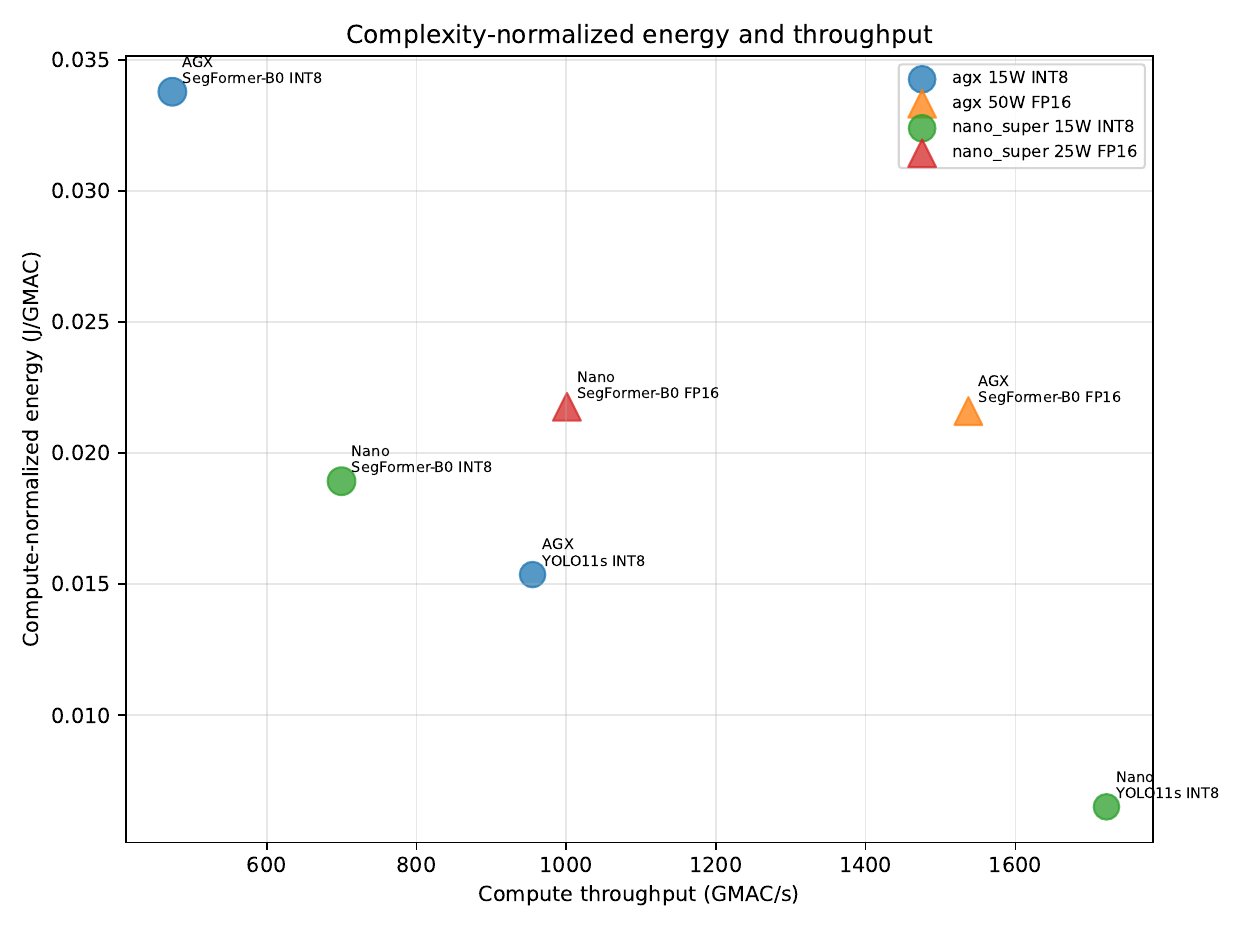}}
    {\IfFileExists{Fig5_j_gmac_vs_gmac_s.png}
    {\includegraphics[width=\linewidth]{Fig5_j_gmac_vs_gmac_s.png}}
    {\fbox{\parbox{0.92\linewidth}{\centering Placeholder for Fig5\_j\_gmac\_vs\_gmac\_s.}}}}
    \caption{Compute-normalized energy view. Matched-budget Nano Super INT8 runs achieved lower J/GMAC than AGX. In native \segformer{} FP16, AGX achieved higher GMAC/s while maintaining nearly the same J/GMAC as Nano Super.}
    \label{fig:jgmac_gmacs}
\end{figure}

Figure~\ref{fig:jgmac_gmacs} shows the complexity-normalized deployment view. The matched 15 W Nano configurations had lower J/GMAC than AGX for both \yoloS{} and \segformer{}. In contrast, native \segformer{} FP16 showed nearly identical J/GMAC for AGX and Nano, while AGX delivered substantially higher GMAC/s.

\subsection{Battery runtime and inference capacity}
\label{sec:results_battery_runtime}

\begin{figure}[t]
    \centering
    \IfFileExists{Fig6_battery_runtime_total_inferences.pdf}
    {\includegraphics[width=\linewidth]{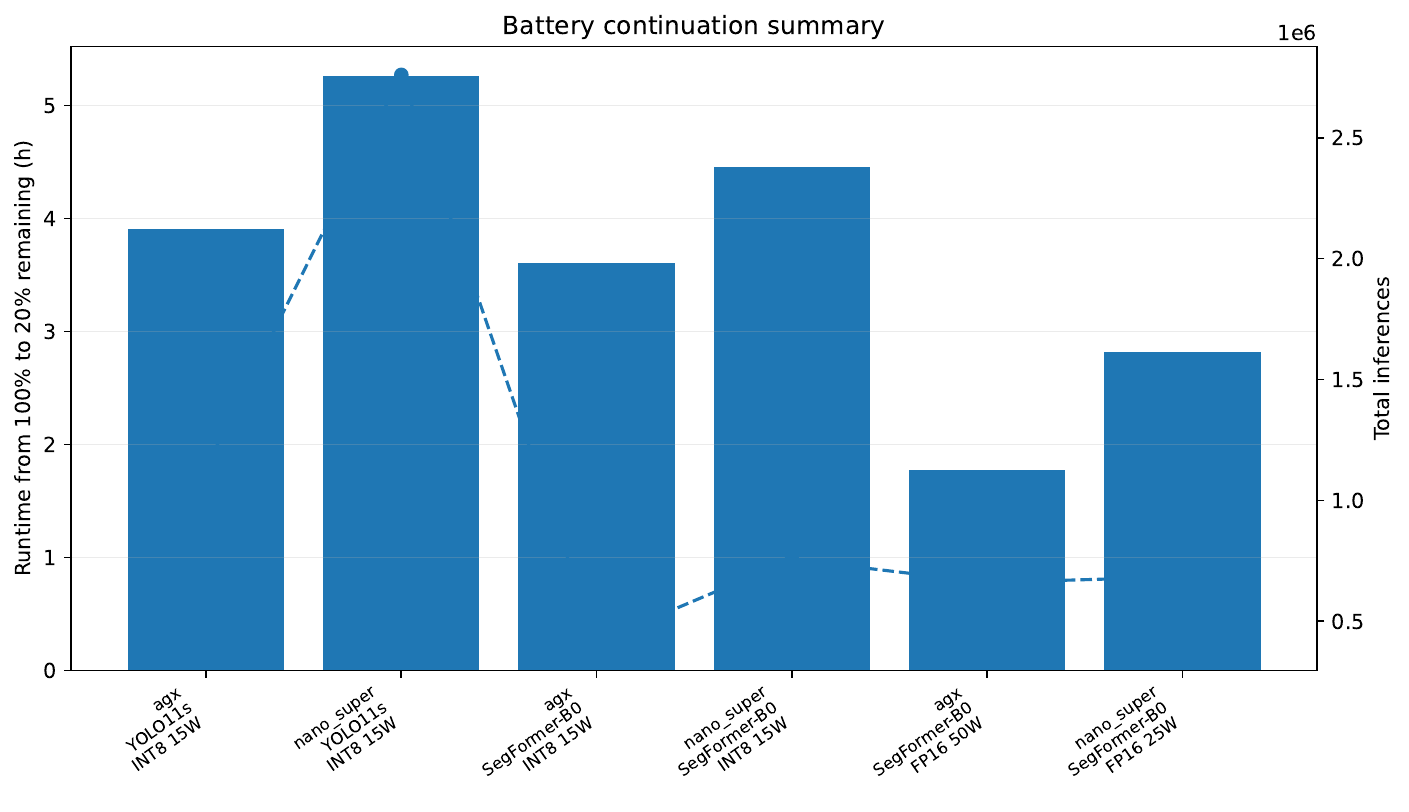}}
    {\IfFileExists{Fig6_battery_runtime_total_inferences.png}
    {\includegraphics[width=\linewidth]{Fig6_battery_runtime_total_inferences.png}}
    {\fbox{\parbox{0.92\linewidth}{\centering Placeholder for Fig6\_battery\_runtime\_total\_inferences.}}}}
    \caption{Battery continuation runtime and total inference count. All main runs used the power bank from 100\% to 20\% remaining and were stopped cleanly. Nano Super ran longer in lower-power matched-budget configurations, while AGX delivered higher native \segformer{} FP16 throughput.}
    \label{fig:battery_runtime}
\end{figure}

Figure~\ref{fig:battery_runtime} summarizes runtime and total inference count over the 100\% to 20\% power-bank continuation window. Nano Super ran longer in matched-budget INT8 configurations because its observed external power was lower. It also processed more total inferences for both \yoloS{} and \segformer{} INT8. In native \segformer{} FP16, Nano ran longer at lower power, but AGX processed inferences at much higher throughput and achieved a similar total inference count over a shorter time.

\subsection{Internal versus external power agreement}
\label{sec:results_agreement}

The time-aligned agreement analysis used 76,691 paired 1 Hz overlap samples from the six main battery runs. These samples were generated by mean-resampling external and internal power to one-second UTC bins and retaining only bins where both channels were present. They were used to quantify agreement and bias, not as the raw external logger sampling rate or as independent experimental replicates. The pooled internal-minus-external bias was $-1.988$ W, with MAE of 2.402 W and RMSE of 2.720 W. The pooled MAPE was 15.65\%. Pearson and Spearman correlations were 0.970 and 0.932, respectively. Thus, internal telemetry followed the external power trend but had non-negligible absolute bias. Because pooled correlations can be inflated by between-run differences in power level, per-run Pearson correlations are reported separately in Table~\ref{tab:power_agreement_results}. Bias, MAE, RMSE, and Bland--Altman limits were therefore treated as the main agreement evidence rather than correlation alone. Because adjacent 1 Hz samples within each run are temporally autocorrelated, the pooled sample count was used to characterize agreement over time and was not treated as independent replication.

Table~\ref{tab:power_agreement_results} summarizes per-run agreement. In matched-budget INT8 runs, internal telemetry was generally below external board-input power by approximately 2.0--2.6 W. The native AGX \segformer{} FP16 run showed a positive internal-minus-external bias of 2.015 W, illustrating that internal rail aggregation can differ by workload and board state. These findings support the measurement hierarchy used in this paper: external inline logging is the primary board-input energy reference, and internal telemetry is a diagnostic channel.

\begin{table*}[t]
\centering
\caption{Time-aligned internal--external power agreement over paired 1 Hz overlap samples. Bias is internal minus external power. \textbf{Sample counts are paired one-second bins used to characterize temporal agreement and were not treated as independent experimental replicates.}}
\label{tab:power_agreement_results}
\scriptsize
\setlength{\tabcolsep}{4pt}
\begin{adjustbox}{max width=\textwidth}
\begin{tabular}{l l l r r r r r r}
\toprule
Board & Mode & Model/precision & Samples & Ext. W & Int. W & Bias W & MAE W & Pearson $r$ \\
\midrule
AGX & 15 W & \yoloS{} INT8 & 13,330 & 14.729 & 12.141 & -2.589 & 2.589 & 0.502 \\
AGX & 15 W & \segformer{} INT8 & 12,337 & 16.229 & 14.150 & -2.079 & 2.085 & 0.585 \\
Nano & 15 W & \yoloS{} INT8 & 18,724 & 11.150 & 9.146 & -2.004 & 2.074 & 0.395 \\
Nano & 15 W & \segformer{} INT8 & 16,029 & 13.145 & 10.974 & -2.171 & 2.231 & 0.575 \\
AGX & 50 W & \segformer{} FP16 & 6,091 & 34.136 & 36.151 & 2.015 & 2.536 & 0.822 \\
Nano & 25 W & \segformer{} FP16 & 10,180 & 21.782 & 18.612 & -3.169 & 3.335 & 0.722 \\
Pooled & Mixed & All main runs & 76,691 & 16.243 & 14.255 & -1.988 & 2.402 & 0.970 \\
\bottomrule
\end{tabular}
\end{adjustbox}
\end{table*}

\begin{figure}[t]
    \centering
    \IfFileExists{Fig7a_internal_external_power_scatter.pdf}
    {\includegraphics[width=\linewidth]{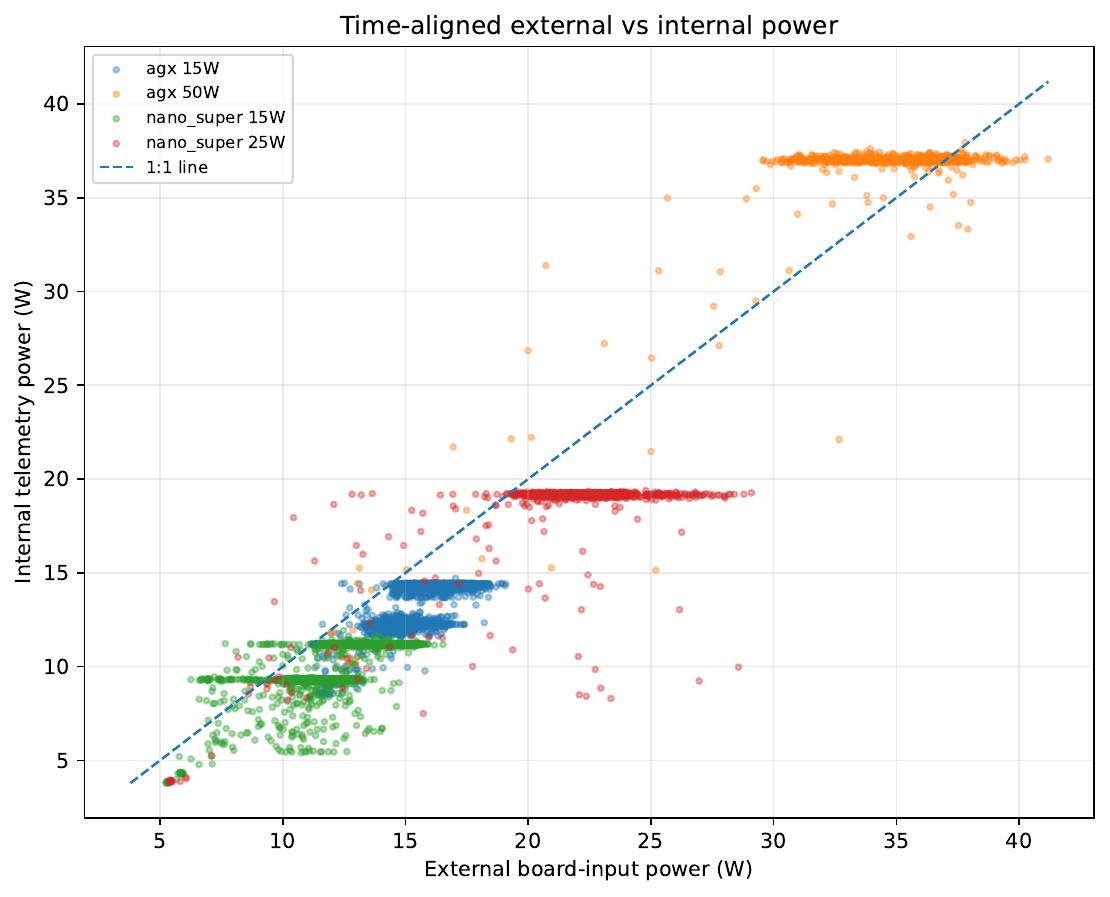}}
    {\IfFileExists{Fig7a_internal_external_power_scatter.png}
    {\includegraphics[width=\linewidth]{Fig7a_internal_external_power_scatter.png}}
    {\fbox{\parbox{0.92\linewidth}{\centering Placeholder for Fig7a\_internal\_external\_power\_scatter.}}}}
    \caption{Time-aligned internal versus external power scatter. Internal telemetry correlates with external board-input power but exhibits workload- and board-dependent bias.}
    \label{fig:internal_external_scatter}
\end{figure}

\begin{figure}[t]
    \centering
    \IfFileExists{Fig7_bland_altman_power_agreement.pdf}
    {\includegraphics[width=\linewidth]{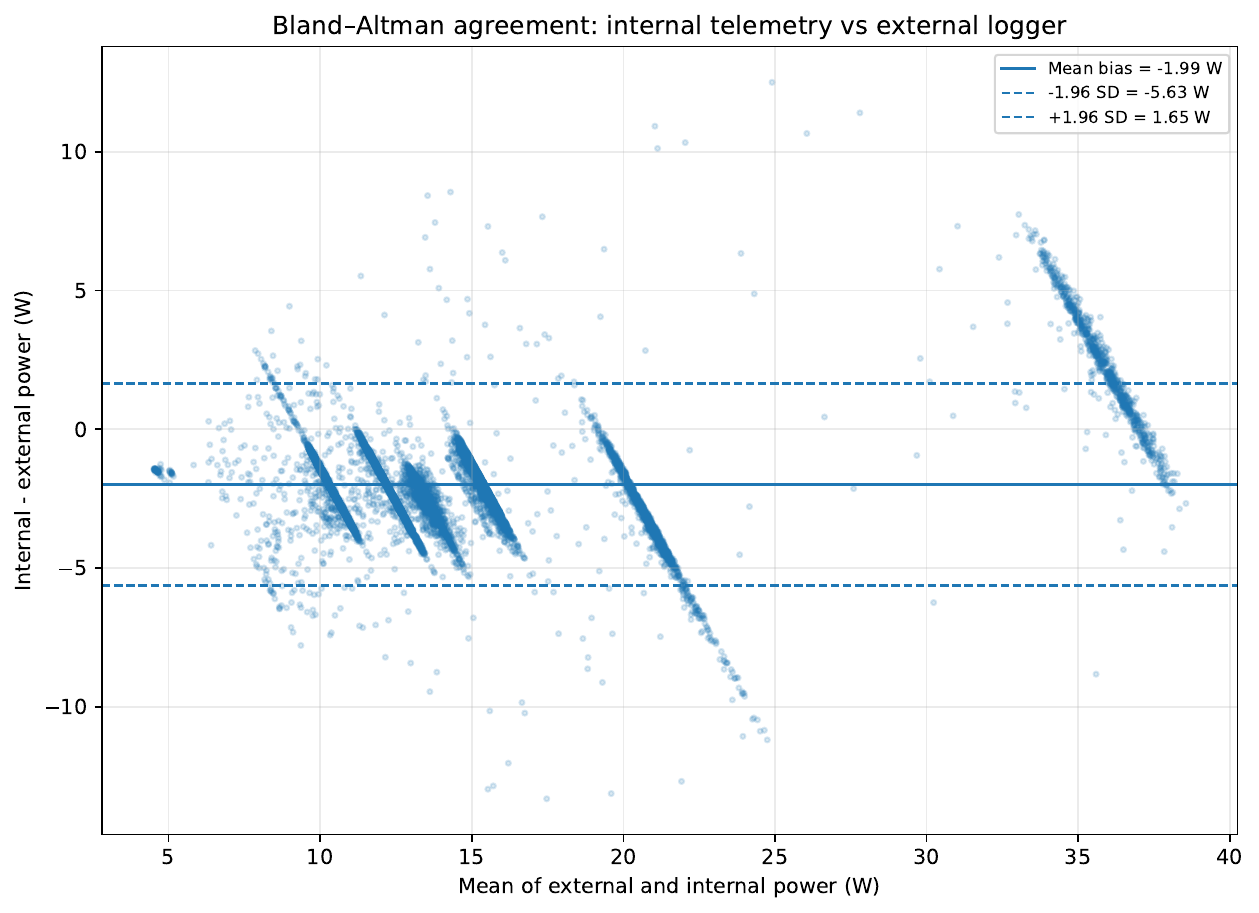}}
    {\IfFileExists{Fig7_bland_altman_power_agreement.png}
    {\includegraphics[width=\linewidth]{Fig7_bland_altman_power_agreement.png}}
    {\fbox{\parbox{0.92\linewidth}{\centering Placeholder for Fig7\_bland\_altman\_power\_agreement.}}}}
    \caption{Bland--Altman analysis of internal and external power. The pooled mean bias was $-1.988$ W, with limits of agreement from approximately $-5.627$ W to $1.651$ W. The observed bias supports using the external inline logger as the primary board-input energy reference.}
    \label{fig:bland_altman}
\end{figure}

\begin{figure}[t]
    \centering
    \IfFileExists{Fig7c_time_aligned_power_trace_example.pdf}
    {\includegraphics[width=\linewidth]{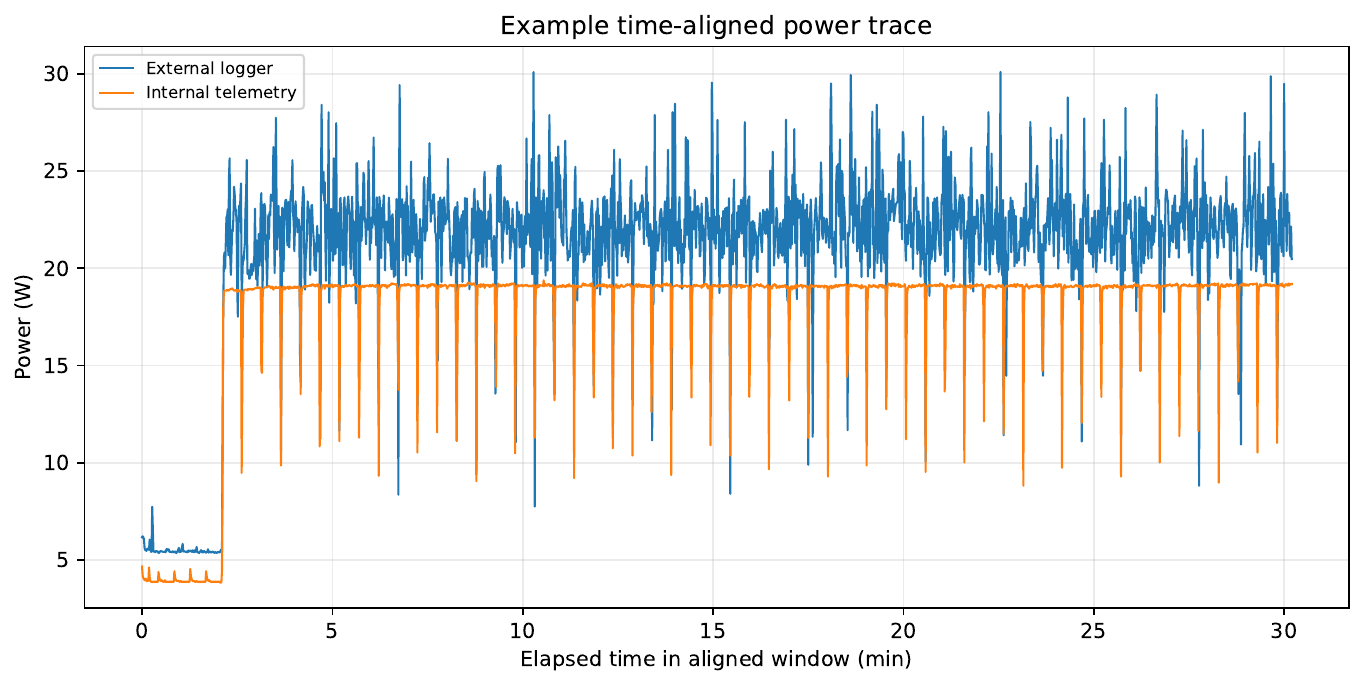}}
    {\IfFileExists{Fig7c_time_aligned_power_trace_example.png}
    {\includegraphics[width=\linewidth]{Fig7c_time_aligned_power_trace_example.png}}
    {\fbox{\parbox{0.92\linewidth}{\centering Placeholder for Fig7c\_time\_aligned\_power\_trace\_example.}}}}
    \caption{Example time-aligned external and internal power traces. External board-input power and internal telemetry follow similar temporal structure but differ in absolute level depending on board and workload.}
    \label{fig:power_trace}
\end{figure}

\subsection{Idle baselines and empirical energy decomposition}
\label{sec:results_energy_model}

Idle baseline runs estimated board/mode baseline external power $P_0$. AGX consumed 10.920 W in 15 W mode and 11.299 W in 50 W mode during idle. Nano Super consumed 5.308 W in 15 W mode and 5.663 W in 25 W mode. Table~\ref{tab:idle_results} lists the measured baseline values.

\begin{table}[t]
\centering
\caption{External idle baseline power used as $P_0$ in the empirical energy model.}
\label{tab:idle_results}
\small
\setlength{\tabcolsep}{4pt}
\begin{tabular}{l r r r}
\toprule
Board/mode & Duration (s) & Avg idle W & Peak idle W \\
\midrule
AGX 15 W & 300 & 10.920 & 15.96 \\
AGX 50 W & 300 & 11.299 & 16.37 \\
Nano 15 W & 300 & 5.308 & 6.78 \\
Nano 25 W & 300 & 5.663 & 7.73 \\
\bottomrule
\end{tabular}
\end{table}

Using the complexity-aware model,
\[
\hat{J}_{\mathrm{inf}} =
\frac{P_0}{\mathrm{FPS}}+\eta C_m,
\]
the fixed platform term explains why low-throughput configurations can have high energy per inference. For example, AGX 15 W \segformer{} INT8 had a relatively high baseline-to-throughput burden because it processed only 31.95 FPS. Its measured energy was 0.5013 J/inference. In contrast, AGX 50 W \segformer{} FP16 processed 103.61 FPS, reducing the amortized baseline contribution and producing 0.3205 J/inference.

Table~\ref{tab:energy_model_results} summarizes the direct energy decomposition for the six main battery rows. Because the main battery set includes only two matched-budget INT8 models per board and one native FP16 model per board, fitted dynamic slopes should be interpreted as exploratory rather than as a universal predictive model. However, the decomposition provides a useful explanation of the observed regime dependence.

\begin{table*}[t]
\centering
\caption{Complexity-aware energy decomposition using idle baseline power. Dynamic power is observed average external power minus idle baseline power.}
\label{tab:energy_model_results}
\scriptsize
\setlength{\tabcolsep}{4pt}
\begin{adjustbox}{max width=\textwidth}
\begin{tabular}{l l l r r r r r r}
\toprule
Board & Mode & Model/precision & FPS & GMAC/s & Avg W & $P_0$ W & Dynamic W & J/GMAC \\
\midrule
AGX & 15 W & \yoloS{} INT8 & 80.90 & 955.13 & 14.54 & 10.92 & 3.62 & 0.01536 \\
Nano & 15 W & \yoloS{} INT8 & 145.82 & 1721.71 & 11.12 & 5.31 & 5.81 & 0.00650 \\
AGX & 15 W & \segformer{} INT8 & 31.95 & 474.02 & 16.01 & 10.92 & 5.09 & 0.03378 \\
Nano & 15 W & \segformer{} INT8 & 47.18 & 699.96 & 13.15 & 5.31 & 7.84 & 0.01892 \\
AGX & 50 W & \segformer{} FP16 & 103.61 & 1537.23 & 33.20 & 11.30 & 21.90 & 0.02160 \\
Nano & 25 W & \segformer{} FP16 & 67.47 & 1001.01 & 21.79 & 5.66 & 16.12 & 0.02177 \\
\bottomrule
\end{tabular}
\end{adjustbox}
\end{table*}

\subsection{Supplementary end-to-end boundary check}
\label{sec:results_end_to_end}

An AGX \yoloS{} INT8 end-to-end image-pipeline battery run was also archived as a boundary check. This run processed 20,595 images at 1.075 FPS, with median latency of 914.758 ms, external energy of 62.213 Wh, and average external power of 11.695 W. The run was not included in the main cross-board energy comparisons because it measured a different, non-optimized image-pipeline boundary rather than pure TensorRT engine execution. The contrast shows why TensorRT J/inference and image-pipeline J/image should be reported separately.

\subsection{Deployment decision map}
\label{sec:results_decision_map}

Table~\ref{tab:deployment_decision_map} summarizes the deployment interpretation. Within the tested configurations, Nano Super was preferred for matched-budget INT8 operation, including \yoloS{} INT8 and \segformer{} INT8. AGX Orin was preferred for native high-load \segformer{} FP16 when throughput and latency were the main priorities. \rtdetr{} remains a contrast case because AGX is faster but Nano is more energy-efficient.

\begin{table*}[t]
\centering
\caption{AgriJetsonBench deployment decision map.}
\label{tab:deployment_decision_map}
\footnotesize
\setlength{\tabcolsep}{4pt}
\begin{tabularx}{\textwidth}{@{}>{\raggedright\arraybackslash}p{0.28\textwidth}
>{\raggedright\arraybackslash}p{0.34\textwidth}
>{\raggedright\arraybackslash}p{0.32\textwidth}@{}}
\toprule
Deployment goal & \textbf{Preferred tested configuration} & Evidence \\
\midrule
Battery-oriented lightweight detection at matched 15 W
& Nano Super 15 W + \yoloS{} INT8 640
& 145.8 FPS; 0.077 J/inference vs AGX 80.9 FPS; 0.181 J/inference \\
Matched-budget INT8 segmentation
& Nano Super 15 W + \segformer{} INT8 640
& 47.2 FPS; 0.281 J/inference vs AGX 31.9 FPS; 0.501 J/inference \\
Native high-load FP16 segmentation throughput
& AGX Orin 50 W + \segformer{} FP16 640
& 103.6 FPS vs Nano 67.5 FPS; lower p95 latency \\
Native high-load FP16 segmentation energy
& Near tie; marginal AGX edge in long battery run
& AGX 0.320 J/inference vs Nano 0.323 J/inference \\
Highest \segformer{} FP16 compute throughput
& AGX Orin 50 W
& 1537 GMAC/s vs Nano 1001 GMAC/s \\
Measurement boundary interpretation
& Pure-engine main; end-to-end supplementary
& Do not mix TensorRT J/inference with image-pipeline J/image \\
General deployment rule
& Choose by workload, precision, power mode, and boundary
& Nano wins matched-budget INT8; AGX wins native high-load FP16 performance \\
\bottomrule
\end{tabularx}
\end{table*}

\subsection{Overall deployment finding}
\label{sec:results_overall}

The final results show that the deployment winner is not fixed by board class. In matched-budget 15 W INT8 operation, Nano Super consistently outperformed AGX Orin for both \yoloS{} and \segformer{}. Nano achieved higher throughput, lower latency, lower external power, lower J/inference, and lower J/GMAC.

In native high-load \segformer{} FP16 operation, AGX Orin became the stronger performance platform. It achieved substantially higher throughput and lower latency than Nano Super while maintaining comparable J/inference and J/GMAC. The native \rtdetr{} FP16 contrast case shows that this is not a universal AGX energy-efficiency win; model architecture and TensorRT execution behaviour also matter.

The concise deployment conclusion is: Nano Super wins matched-budget INT8; AGX Orin wins native high-load FP16 performance. More generally, agricultural edge-AI deployment efficiency is governed jointly by model complexity, precision, TensorRT execution boundary, batch size, and Jetson power mode, rather than by board class alone.

%% file: discussion.tex

\section{Discussion}
\label{sec:discussion}

\subsection{Principal findings}
\label{sec:discussion_principal_findings}

\bench{} should be interpreted as a deployment-measurement benchmark rather than a new model-architecture study. Its main finding is that agricultural edge-AI efficiency cannot be inferred from validation accuracy, board class, or model size alone. The preferred Jetson platform changed with workload, precision, timing boundary, and operating mode: in the \matched{} 15 W INT8 regime, \nano{} outperformed \agx{} for both \yoloS{} and \segformer{}, whereas in the \nativeview{} \segformer{} FP16 regime, \agx{} delivered substantially higher throughput and lower latency with comparable energy per inference and compute-normalized energy.

This regime dependence supports the central premise of \bench{}: deployment decisions for precision-agriculture vision systems require joint reporting of task accuracy, latency, tail latency, board-input power, energy per inference, thermal stability, model complexity, and measurement boundary \cite{liakos2018machine,kamilaris2018deep,shamshiri2018research}. Separating the \matched{} and \nativeview{} comparisons is therefore essential because they answer different engineering questions and can produce different board rankings.

\subsection{Accuracy-only model selection is insufficient}
\label{sec:discussion_accuracy_only}

The training results show that all retained models were valid deployment candidates, but the most accurate or most computationally expensive model was not automatically the best edge-AI choice. \segformer{} achieved the strongest segmentation accuracy, whereas \deeplab{} required far more computation and a much larger artifact footprint. Similarly, \rtdetr{} required more GMACs than \yoloS{} but achieved lower detection test mAP50-95. This pattern is consistent with the broader deployment challenge in agricultural computer vision: accuracy-focused model development and resource-constrained field inference are related but not identical objectives \cite{yoloweeds2023,aicropcam2023,islam2025weed}.

The complexity-versus-accuracy result in Fig.~\ref{fig:complexity_accuracy} therefore motivates the deployment measurements rather than replacing them. Parameter count and GMACs are useful explanatory variables, but they do not fully determine either accuracy or deployment cost. Measured TensorRT latency, external power, and energy-normalized metrics remain necessary when comparing heterogeneous model families such as YOLO detectors, transformer-style detectors, lightweight CNN segmenters, and transformer-style segmentation models \cite{ultralytics_yolo11,lv2024rtdetr,yu2021bisenetv2,howard2019mobilenetv3,chen2018deeplabv3plus,xie2021segformer}.

\subsection{Why Nano Super wins matched-budget INT8 inference}
\label{sec:discussion_nano_matched}

The clearest result in the \matched{} setting was the \nano{} advantage for INT8 inference. For \yoloS{} INT8 at 15 W, \nano{} achieved higher throughput, lower latency, lower average external power, and lower energy per inference than \agx{}, reducing energy from 0.1814 to 0.0767 J/inference. The same pattern held for \segformer{} INT8 at 15 W, showing that the matched-budget Nano advantage was not limited to a lightweight detector but also applied to a transformer-style segmentation workload.

A likely explanation is that \agx{} is underutilized in this low-power, batch-1 INT8 regime. The larger platform has more available compute capacity in native modes, but at 15 W it cannot expose that capacity effectively for these workloads and retains higher platform overhead than \nano{}. Jetson power-mode policies can constrain clocks, rail behaviour, and available compute resources \cite{nvidia_jetson_orin,nvidia_jetson_power}. The idle baselines support this interpretation: the AGX idle baseline was approximately 10.9 W in 15 W mode, whereas the Nano idle baseline was approximately 5.3 W. When throughput is low, this fixed overhead is amortized across fewer inferences, increasing J/inference.

\subsection{Why AGX Orin becomes favourable in native SegFormer-B0 FP16}
\label{sec:discussion_agx_native}

The native \segformer{} FP16 result shows the other side of the deployment trade-off. In 50 W native/high-performance mode, \agx{} reached 103.61 FPS for \segformer{} FP16, compared with 67.47 FPS on \nano{} in 25 W mode, and it reduced median and p95 latency substantially. Importantly, this higher throughput did not produce a meaningful energy-per-inference penalty: AGX required 0.3205 J/inference, while Nano required 0.3229 J/inference in the inference-aligned long battery comparison, with nearly tied compute-normalized energy.

This result does not imply that \agx{} is always more energy-efficient than \nano{} or that higher power modes always improve efficiency. It shows that a larger platform can become favourable when the workload is heavy enough to expose its compute capacity and when throughput is important. TensorRT optimization, precision mode, kernel selection, and engine construction can strongly affect measured inference behaviour on embedded GPUs \cite{nvidia_tensorrt,nvidia_trtexec}. Thus, battery-oriented lightweight detection favours \nano{} in the tested matched-budget INT8 regime, whereas low-latency heavier segmentation can justify \agx{} in native mode.

\subsection{RT-DETR-R18 shows that architecture still matters}
\label{sec:discussion_rtdetr}

The \rtdetr{} FP16 native result prevents an overly simple conclusion such as ``AGX wins native mode'' or ``Nano wins energy.'' AGX was faster for \rtdetr{}, but Nano remained more energy-efficient. Architecture-specific TensorRT behaviour, memory access patterns, precision handling, and kernel fusion can therefore change the energy ranking, even when a larger board improves throughput \cite{lv2024rtdetr,nvidia_tensorrt}. A short, well-instrumented pure-engine benchmark remains necessary for each architecture and precision mode.

\subsection{Importance of timing-boundary separation}
\label{sec:discussion_boundary}

One of the main methodological lessons from this study is that timing boundary must be explicit. The main results use a pure TensorRT engine-level boundary to isolate accelerator-side model execution, precision mode, and power-mode policy. This boundary enables fair comparison between random-input \texttt{trtexec} style execution and preloaded tensor execution when the timed region contains only batch-1 TensorRT inference \cite{nvidia_tensorrt,nvidia_trtexec}.

Pure-engine results should not be interpreted as full application-level image-pipeline throughput. The archived AGX \yoloS{} INT8 end-to-end image-pipeline run was much slower because it included image-path handling, preprocessing, postprocessing, and saving overhead. That supplementary run illustrates the possible gap between engine-level inference and a full application pipeline, but it was not mixed with the main comparison to avoid confounding model execution efficiency with pipeline implementation overhead. For agricultural deployment, pure-engine results answer how efficient the optimized TensorRT model is on the Jetson platform, while end-to-end results answer how efficient a specific application implementation is \cite{aicropcam2023,rtal2023,tinysegformer2024,islam2025weed}.

\subsection{External board-input power should remain the primary energy reference}
\label{sec:discussion_power_measurement}

The time-aligned internal--external agreement analysis showed strong temporal correlation but non-negligible absolute bias. Using the paired 1 Hz overlap workflow described in Section~\ref{sec:agreement_methods}, 76,691 paired samples were compared. Internal telemetry and external power were strongly correlated, but the pooled internal-minus-external bias was approximately -1.99 W, with MAE of 2.40 W and RMSE of 2.72 W. The Bland--Altman limits of agreement further showed that internal telemetry can deviate from board-input power by several watts depending on board, mode, and workload \cite{bland1986statistical}.

These results support the measurement hierarchy used throughout \bench{}. The external inline logger should be treated as the primary board-input energy reference, while internal telemetry should be used as a diagnostic channel for temperature, memory pressure, rail-level behaviour, temporal power trends, and abnormal run conditions. Because rail aggregation and coverage differ across devices and operating states, internal telemetry should not replace external board-input logging when reporting energy per inference \cite{nvidia_tegrastats,nvidia_jetson_power,ti_ina260}.

\subsection{Battery continuation adds deployment realism}
\label{sec:discussion_battery}

Short benchmark runs are useful for latency and throughput characterization, but they do not fully represent sustained field operation. The power-bank continuation protocol added a deployment-oriented layer by running from 100\% power-bank charge to 20\% remaining and stopping cleanly before cutoff. Nano Super ran longer in the matched-budget INT8 configurations and processed more total inferences for both \yoloS{} and \segformer{} INT8. In native \segformer{} FP16, Nano ran longer because it used lower power, but AGX processed at a higher rate and achieved a similar total inference count over a shorter period.

The battery protocol also made the stop condition explicit. The power-bank display was used only to define the 100\% to 20\% continuation window; all energy metrics were computed from the external logger. This distinction is important because power-source state-of-charge indicators are not equivalent to time-integrated board-input energy.

\subsection{Interpretation of the empirical energy model}
\label{sec:discussion_energy_model}

The complexity-aware energy model was used to interpret the observed regime dependence, not to claim universal predictive accuracy. The model separates measured external power into a board/mode baseline term and a workload-dependent compute term:
\begin{equation}
\hat{P}_{\mathrm{ext}} = P_0 + \eta (C_m \times \mathrm{FPS}),
\end{equation}
with corresponding energy per inference:
\begin{equation}
\hat{J}_{\mathrm{inf}} =
\frac{P_0}{\mathrm{FPS}} + \eta C_m.
\end{equation}

This form explains why a larger board may appear inefficient under a low-power batch-1 regime: if $P_0$ is high and FPS is low, the fixed overhead per inference becomes large. Conversely, when AGX was operated in 50 W mode with \segformer{} FP16, throughput increased substantially, reducing the amortized baseline contribution and producing energy per inference comparable to Nano. Because the main battery set contains six long continuation runs, the fitted dynamic slopes should be interpreted as exploratory deployment estimators rather than universal coefficients.

\subsection{Implications for agricultural edge-AI deployment}
\label{sec:discussion_practical}

The results lead to four practical recommendations. First, for battery-oriented lightweight detection at matched 15 W, \nano{} with \yoloS{} INT8 is the preferred configuration among those tested. Second, for matched-budget \segformer{} INT8 segmentation, \nano{} again provides better throughput and energy efficiency than \agx{}. Third, for native high-load \segformer{} FP16 segmentation, \agx{} is the preferred performance platform when latency or frame rate is more important than maximum runtime duration. Fourth, model architecture and precision mode must be evaluated directly; the \rtdetr{} contrast case shows that a native-mode speed advantage does not automatically imply an energy advantage \cite{yoloweeds2023,aicropcam2023,rtal2023,tinysegformer2024,islam2025weed,wang2026livestock}.

Pure-engine and end-to-end results should also be reported separately. A user deploying a full field pipeline should expect additional overhead from image capture, decoding, preprocessing, postprocessing, storage, and communication. The deployment rule emerging from \bench{} is therefore:
\begin{quote}
Choose the board, precision, and power mode according to workload complexity, latency target, energy budget, and measurement boundary. Nano Super wins matched-budget INT8 in this study, while AGX Orin wins native high-load FP16 segmentation performance.
\end{quote}

\subsection{Reproducibility and auditability}
\label{sec:discussion_reproducibility}

A major goal of \bench{} is to make the benchmark auditable. The package records dataset splits, class names, license notes, training summaries, validation metrics, model artifact manifests, SHA256 checksums, TensorRT deployment evidence, hardware/software versions, battery summaries, latency CSV files, event records, noise-guard summaries, sanitized run-validity summaries, external logger archives, internal telemetry, idle baselines, agreement tables, and analysis scripts. This structure allows the reported tables to be checked against run-level evidence.

This matters because edge-AI benchmarking is vulnerable to ambiguity. FPS and energy values are difficult to interpret without knowing the input size, batch size, timing boundary, precision mode, power mode, and whether energy came from internal telemetry, external board-input logging, or a battery-percentage estimate. By preserving metadata, code, manifests, checksums, and raw evidence, \bench{} reduces this ambiguity and provides a starting point for larger agricultural edge-AI benchmark extensions \cite{wilkinson2016fair,peng2011reproducible}.

\subsection{Limitations}
\label{sec:discussion_limitations}

Several limitations should be considered when interpreting the results. First, the main Jetson deployment metrics use a pure TensorRT engine-level boundary. This was intentional for controlled comparison, but it does not measure full application-level image processing, which depends on camera input, decoding, preprocessing, postprocessing, storage, and communication. Second, all main runs used batch size 1. This reflects online agricultural inference, but it may underutilize larger platforms such as AGX Orin; future work could include batch-size scaling while keeping batch-1 as the main deployment baseline.

Third, the long battery continuation tests were designed as sustained deployment trials rather than statistical repeat experiments. Short benchmark repeats and validity checks support run quality, but repeated long battery trials are needed to quantify run-to-run variability. Fourth, the empirical energy model is useful for interpretation but should not be treated as universal across all Jetson boards, workloads, and environmental conditions. Fifth, the evaluated datasets represent crop/weed detection and segmentation, so generalization to livestock monitoring, fruit counting, disease detection, or autonomous navigation should be tested directly.

Finally, the main 640 cross-board INT8 engines were accompanied by retention evidence relative to board-matched FP16 TensorRT references, and all four main engines passed the aggregate retention criteria used in this manuscript. This supports the matched-budget INT8 deployment comparisons reported for YOLO11s and SegFormer-B0. However, the retention check was scoped to the main 640 engines and should not be interpreted as a complete retention guarantee for every supplementary INT8 engine, input resolution, or model family.

\subsection{Future work}
\label{sec:discussion_future}

Future work should extend \bench{} to additional agricultural tasks and datasets, including fruit detection, plant disease segmentation, livestock monitoring, and navigation-relevant perception. More edge platforms should be added, including other Jetson Orin variants and non-NVIDIA embedded accelerators, while preserving the same external power and validity protocol. A useful supplementary runtime-level comparison would evaluate ONNX Runtime with the TensorRT Execution Provider, which would address deployments that prefer the ONNX Runtime API but should be reported separately from the pure TensorRT engine-level boundary used as the main benchmark in this study \citep{onnxruntime_tensorrt_ep}. Full end-to-end application pipelines should also be benchmarked separately from pure-engine inference, and the empirical energy model should be expanded with more models, repeated runs, input sizes, and batch sizes. The next step is field validation of the most promising configurations on mobile platforms with real sensors, enclosure constraints, variable ambient temperatures, and realistic duty cycles.

%% file: conclusions.tex

\section{Conclusions}
\label{sec:conclusions}

This study introduced \bench{}, a reproducible complexity-aware energy benchmark for agricultural vision model deployment on NVIDIA Jetson edge-AI platforms. The benchmark connects locked agricultural datasets, offline model training, ONNX and TensorRT deployment, pure-engine Jetson benchmarking, external board-input power measurement, internal telemetry, battery continuation tests, and deployment decision analysis. By preserving tables, figures, scripts, telemetry, logger archives, manifests, and SHA256 checksums, \bench{} was designed not only to report results, but also to make the evidence behind those results auditable.

The results show that agricultural edge-AI deployment efficiency is not determined by board class alone. In the matched-budget 15 W INT8 regime, \nano{} was the preferred platform for the tested workloads. For \yoloS{} INT8, Nano Super achieved higher throughput, lower latency, and lower energy per inference than \agx{}. For \segformer{} INT8, Nano Super again achieved higher FPS and lower J/inference under the same nominal 15 W budget. These results support Nano Super as the better choice for battery-oriented matched-budget INT8 deployment in the evaluated detection and segmentation workloads.

The native/high-performance \segformer{} FP16 comparison showed a different deployment regime. When \agx{} was operated in 50 W mode and Nano Super in 25 W mode, AGX Orin achieved substantially higher throughput and lower latency for \segformer{} FP16 while maintaining nearly comparable energy per inference and compute-normalized energy. This result shows that the larger AGX platform can become favourable when the workload is sufficiently heavy and the deployment objective emphasizes latency or throughput rather than minimum average power. The \rtdetr{} FP16 contrast case further showed that architecture matters: AGX can be faster while Nano remains more energy-efficient for some models.

The dual-source power analysis also supports a clear measurement recommendation. Internal Jetson telemetry is useful for diagnostics, temperature tracking, memory and rail-level behaviour, and time-aligned agreement analysis. However, the observed internal--external bias confirms that external inline board-input logging should remain the primary reference for reporting energy per inference. For agricultural edge-AI studies, this distinction is important because internal telemetry, external board-input energy, and battery state-of-charge indicators are not interchangeable.

The study also emphasizes the need to separate measurement boundaries. Pure TensorRT engine-level inference provides a controlled way to compare models, precisions, power modes, and Jetson platforms. End-to-end image-pipeline measurements are useful for complete application profiling, but they should not be mixed with pure-engine J/inference results. This boundary separation is essential for interpretable energy-normalized benchmarking.

Overall, the main deployment rule from \bench{} is:

\begin{quote}
Nano Super wins matched-budget INT8 inference in the tested workloads, while AGX Orin wins native high-load \segformer{} FP16 performance. Board choice should therefore be made according to workload complexity, precision mode, power mode, latency target, energy budget, and measurement boundary.
\end{quote}

These findings provide a practical framework for selecting agricultural edge-AI deployments. Rather than relying on accuracy alone, future field systems should evaluate model accuracy together with latency, external board-input power, energy per inference, compute-normalized energy, thermal behaviour, and run validity. Future work should extend \bench{} to additional agricultural tasks, edge accelerators, repeated long-duration field trials, batch-size scaling studies, and fully optimized end-to-end application pipelines.

%% file: availability.tex
\section*{Data and code availability}

The data and code that support the findings of this study are available from the corresponding author upon reasonable request.

%% file: declarations.tex

\section*{Declaration of competing interest}

The authors declare that they have no known competing financial interests or personal relationships that could have appeared to influence the work reported in this paper.

\section*{Funding}

This research received no external funding.

\section*{CRediT authorship contribution statement}

Hasan Jahanifar: Conceptualization, Methodology, Software, Data curation, Investigation, Formal analysis, Visualization, Writing -- original draft. Hasan Mirzakhaninafchi: Methodology, Validation, Writing -- review and editing. Wesley M. Porter: Supervision, Resources, Project administration, Validation, Writing -- review and editing. Abolfazl Najar: Methodology, Investigation, Validation. Glen C. Rains: Supervision, Resources, Validation, Writing -- review and editing.

%% file: references.bib
@article{liakos2018machine,
  author  = {Liakos, Konstantinos G. and Busato, Patrizia and Moshou, Dimitrios and Pearson, Simon and Bochtis, Dionysis},
  title   = {Machine Learning in Agriculture: A Review},
  journal = {Sensors},
  volume  = {18},
  number  = {8},
  pages   = {2674},
  year    = {2018},
  doi     = {10.3390/s18082674},
  url     = {https://doi.org/10.3390/s18082674}
}

@article{kamilaris2018deep,
  author  = {Kamilaris, Andreas and Prenafeta-Bold{\'u}, Francesc X.},
  title   = {Deep learning in agriculture: A survey},
  journal = {Computers and Electronics in Agriculture},
  volume  = {147},
  pages   = {70--90},
  year    = {2018},
  doi     = {10.1016/j.compag.2018.02.016},
  url     = {https://doi.org/10.1016/j.compag.2018.02.016}
}

@article{shamshiri2018research,
  author  = {Shamshiri, Redmond Ramin and Weltzien, Cornelia and Hameed, Ibrahim A. and Yule, Ian J. and Grift, Tony E. and Balasundram, Siva K. and Pitonakova, Lenka and Ahmad, Desa and Chowdhary, Girish},
  title   = {Research and development in agricultural robotics: A perspective of digital farming},
  journal = {International Journal of Agricultural and Biological Engineering},
  volume  = {11},
  number  = {4},
  pages   = {1--14},
  year    = {2018},
  doi     = {10.25165/j.ijabe.20181104.4278},
  url     = {https://doi.org/10.25165/j.ijabe.20181104.4278}
}

@article{yoloweeds2023,
  author  = {Dang, Fengying and Chen, Dong and Lu, Yuzhen and Li, Zhaojian},
  title   = {{YOLOWeeds}: A novel benchmark of {YOLO} object detectors for multi-class weed detection in cotton production systems},
  journal = {Computers and Electronics in Agriculture},
  volume  = {205},
  pages   = {107655},
  year    = {2023},
  doi     = {10.1016/j.compag.2023.107655},
  url     = {https://doi.org/10.1016/j.compag.2023.107655}
}

@article{aicropcam2023,
  author  = {Chamara, Nipuna and Bai, Geng and Ge, Yufeng},
  title   = {{AICropCAM}: Deploying classification, segmentation, detection, and counting deep-learning models for crop monitoring on the edge},
  journal = {Computers and Electronics in Agriculture},
  volume  = {215},
  pages   = {108420},
  year    = {2023},
  doi     = {10.1016/j.compag.2023.108420},
  url     = {https://doi.org/10.1016/j.compag.2023.108420}
}

@article{rtal2023,
  author  = {Gao, Rui and Chang, Penghao and Chang, Dong and Tian, Xin and Li, Yan and Ruan, Zhiwen and Su, Zhongbin},
  title   = {{RTAL}: An edge computing method for real-time rice lodging assessment},
  journal = {Computers and Electronics in Agriculture},
  volume  = {215},
  pages   = {108386},
  year    = {2023},
  doi     = {10.1016/j.compag.2023.108386},
  url     = {https://doi.org/10.1016/j.compag.2023.108386}
}

@article{tinysegformer2024,
  author  = {Zhang, Yan and Lv, Chunli},
  title   = {{TinySegformer}: A lightweight visual segmentation model for real-time agricultural pest detection},
  journal = {Computers and Electronics in Agriculture},
  volume  = {218},
  pages   = {108740},
  year    = {2024},
  doi     = {10.1016/j.compag.2024.108740},
  url     = {https://doi.org/10.1016/j.compag.2024.108740}
}

@article{islam2025weed,
  author  = {Islam, Md Didarul and Liu, Wenxin and Izere, Pascal and Singh, Puranjit and Yu, Chun and Riggan, Benjamin and Zhang, Kai and Jhala, Amit J. and Knezevic, Stevan and Ge, Yufeng and Pitla, Santosh and Luck, Joe and Shi, Yeyin},
  title   = {Towards real-time weed detection and segmentation with lightweight {CNN} models on edge devices},
  journal = {Computers and Electronics in Agriculture},
  volume  = {237},
  pages   = {110600},
  year    = {2025},
  doi     = {10.1016/j.compag.2025.110600},
  url     = {https://doi.org/10.1016/j.compag.2025.110600}
}

@article{wang2026livestock,
  author  = {Wang, Yiwei and Yuan, Xufeng and Wei, Bingxue and Ruchay, Alexey and Pezzuolo, Andrea and Guo, Hao},
  title   = {Performance evaluation of a state-of-the-art keypoint detection method for precision livestock farming},
  journal = {Computers and Electronics in Agriculture},
  volume  = {240},
  pages   = {111230},
  year    = {2026},
  doi     = {10.1016/j.compag.2025.111230},
  url     = {https://doi.org/10.1016/j.compag.2025.111230}
}

@misc{cottonweeddet12_dataset,
  author       = {{Weed-AI}},
  title        = {{CottonWeedDet12 Dataset}},
  year         = {2023},
  howpublished = {\url{https://weed-ai.sydney.edu.au/datasets/2c14915b-0827-4b65-9908-d2a6df0d48f3}},
}

@inproceedings{cropandweed_dataset,
  author    = {Steininger, Daniel and Trondl, Andreas and Croonen, Gerardus and Simon, Julia and Widhalm, Verena},
  title     = {The {CropAndWeed} Dataset: A Multi-Modal Learning Approach for Efficient Crop and Weed Manipulation},
  booktitle = {Proceedings of the IEEE/CVF Winter Conference on Applications of Computer Vision Workshops},
  pages     = {3729--3738},
  year      = {2023},
  doi       = {10.1109/WACV56688.2023.00372},
  url       = {https://doi.org/10.1109/WACV56688.2023.00372}
}

@misc{ultralytics_yolo11,
  author       = {{Ultralytics}},
  title        = {{Ultralytics YOLO11}},
  year         = {2024},
  howpublished = {\url{https://docs.ultralytics.com/models/yolo11/}},
}

@inproceedings{lv2024rtdetr,
  author    = {Zhao, Yian and Lv, Wenyu and Xu, Shangliang and Wei, Jinman and Wang, Guanzhong and Dang, Qingqing and Liu, Yi and Chen, Jie},
  title     = {{DETRs} Beat {YOLOs} on Real-time Object Detection},
  booktitle = {Proceedings of the IEEE/CVF Conference on Computer Vision and Pattern Recognition},
  pages     = {16965--16974},
  year      = {2024},
  doi       = {10.1109/CVPR52733.2024.01605},
  url       = {https://doi.org/10.1109/CVPR52733.2024.01605}
}

@article{yu2021bisenetv2,
  author  = {Yu, Changqian and Gao, Changxin and Wang, Jingbo and Yu, Gang and Shen, Chunhua and Sang, Nong},
  title   = {{BiSeNet V2}: Bilateral Network with Guided Aggregation for Real-Time Semantic Segmentation},
  journal = {International Journal of Computer Vision},
  volume  = {129},
  pages   = {3051--3068},
  year    = {2021},
  doi     = {10.1007/s11263-021-01515-2},
  url     = {https://doi.org/10.1007/s11263-021-01515-2}
}

@inproceedings{howard2019mobilenetv3,
  author    = {Howard, Andrew and Sandler, Mark and Chu, Grace and Chen, Liang-Chieh and Chen, Bo and Tan, Mingxing and Wang, Weijun and Zhu, Yukun and Pang, Ruoming and Vasudevan, Vijay and Le, Quoc V. and Adam, Hartwig},
  title     = {Searching for {MobileNetV3}},
  booktitle = {Proceedings of the IEEE/CVF International Conference on Computer Vision},
  pages     = {1314--1324},
  year      = {2019},
  doi       = {10.1109/ICCV.2019.00140},
  url       = {https://doi.org/10.1109/ICCV.2019.00140}
}

@inproceedings{chen2018deeplabv3plus,
  author    = {Chen, Liang-Chieh and Zhu, Yukun and Papandreou, George and Schroff, Florian and Adam, Hartwig},
  title     = {Encoder-Decoder with Atrous Separable Convolution for Semantic Image Segmentation},
  booktitle = {Proceedings of the European Conference on Computer Vision},
  pages     = {801--818},
  year      = {2018},
  doi       = {10.1007/978-3-030-01234-2_49},
  url       = {https://doi.org/10.1007/978-3-030-01234-2_49}
}

@inproceedings{xie2021segformer,
  author    = {Xie, Enze and Wang, Wenhai and Yu, Zhiding and Anandkumar, Anima and Alvarez, Jose M. and Luo, Ping},
  title     = {{SegFormer}: Simple and Efficient Design for Semantic Segmentation with Transformers},
  booktitle = {Advances in Neural Information Processing Systems},
  volume    = {34},
  pages     = {12077--12090},
  year      = {2021},
  url       = {https://proceedings.neurips.cc/paper/2021/hash/64f1f27bf1b4ec22924fd0acb550c235-Abstract.html}
}

@inproceedings{akiba2019optuna,
  author    = {Akiba, Takuya and Sano, Shotaro and Yanase, Toshihiko and Ohta, Takeru and Koyama, Masanori},
  title     = {{Optuna}: A Next-generation Hyperparameter Optimization Framework},
  booktitle = {Proceedings of the 25th ACM SIGKDD International Conference on Knowledge Discovery and Data Mining},
  pages     = {2623--2631},
  year      = {2019},
  doi       = {10.1145/3292500.3330701},
  url       = {https://doi.org/10.1145/3292500.3330701}
}

@misc{nvidia_jetson_orin,
  author       = {{NVIDIA}},
  title        = {{NVIDIA Jetson Linux Developer Guide}},
  year         = {2024},
  howpublished = {\url{https://docs.nvidia.com/jetson/archives/r36.4/DeveloperGuide/index.html}},
}

@misc{nvidia_tensorrt,
  author       = {{NVIDIA}},
  title        = {{NVIDIA TensorRT Documentation}},
  year         = {2026},
  howpublished = {\url{https://docs.nvidia.com/deeplearning/tensorrt/latest/index.html}},
}

@misc{nvidia_trtexec,
  author       = {{NVIDIA}},
  title        = {Performance Benchmarking using \texttt{trtexec}},
  year         = {2026},
  howpublished = {\url{https://docs.nvidia.com/deeplearning/tensorrt/latest/performance/benchmarking.html}},
}

@misc{nvidia_tegrastats,
  author       = {{NVIDIA}},
  title        = {{Tegrastats Utility --- NVIDIA Jetson Linux Developer Guide}},
  year         = {2026},
  howpublished = {\url{https://docs.nvidia.com/jetson/archives/r36.5/DeveloperGuide/AT/JetsonLinuxDevelopmentTools/TegrastatsUtility.html}},
}

@misc{nvidia_jetson_power,
  author       = {{NVIDIA}},
  title        = {{Jetson Orin NX Series and Jetson AGX Orin Series: Platform Power and Performance}},
  year         = {2024},
  howpublished = {\url{https://docs.nvidia.com/jetson/archives/r35.1/DeveloperGuide/text/SD/PlatformPowerAndPerformance/JetsonOrinNxSeriesAndJetsonAgxOrinSeries.html}},
}

@misc{ti_ina260,
  author       = {{Texas Instruments}},
  title        = {{INA260 36-V, 16-bit, Precision I2C Output Current/Voltage/Power Monitor with Integrated Shunt Resistor}},
  year         = {2026},
  howpublished = {\url{https://www.ti.com/product/INA260}},
}

@article{bland1986statistical,
  author  = {Bland, J. Martin and Altman, Douglas G.},
  title   = {Statistical methods for assessing agreement between two methods of clinical measurement},
  journal = {The Lancet},
  volume  = {327},
  number  = {8476},
  pages   = {307--310},
  year    = {1986},
  doi     = {10.1016/S0140-6736(86)90837-8},
  url     = {https://doi.org/10.1016/S0140-6736(86)90837-8}
}

@article{wilkinson2016fair,
  author  = {Wilkinson, Mark D. and Dumontier, Michel and Aalbersberg, IJsbrand Jan and Appleton, Gabrielle and Axton, Myles and Baak, Arie and Blomberg, Niklas and Boiten, Jan-Willem and da Silva Santos, Luiz Bonino and Bourne, Philip E. and Bouwman, Jildau and Brookes, Anthony J. and Clark, Tim and Crosas, Merc{\`e} and Dillo, Ingrid and Dumon, Olivier and Edmunds, Scott and Evelo, Chris T. and Finkers, Richard and Gonzalez-Beltran, Alejandra and Gray, Alasdair J. G. and Groth, Paul and Goble, Carole and Grethe, Jeffrey S. and Heringa, Jaap and 't Hoen, Peter A. C. and Hooft, Rob and Kuhn, Tobias and Kok, Ruben and Kok, Joost and Lusher, Scott J. and Martone, Maryann E. and Mons, Albert and Packer, Abel L. and Persson, Bengt and Rocca-Serra, Philippe and Roos, Marco and van Schaik, Rene and Sansone, Susanna-Assunta and Schultes, Erik and Sengstag, Thierry and Slater, Ted and Strawn, George and Swertz, Morris A. and Thompson, Mark and van der Lei, Johan and van Mulligen, Erik and Velterop, Jan and Waagmeester, Andra and Wittenburg, Peter and Wolstencroft, Katherine and Zhao, Jun and Mons, Barend},
  title   = {The {FAIR} Guiding Principles for scientific data management and stewardship},
  journal = {Scientific Data},
  volume  = {3},
  pages   = {160018},
  year    = {2016},
  doi     = {10.1038/sdata.2016.18},
  url     = {https://doi.org/10.1038/sdata.2016.18}
}

@article{peng2011reproducible,
  author  = {Peng, Roger D.},
  title   = {Reproducible Research in Computational Science},
  journal = {Science},
  volume  = {334},
  number  = {6060},
  pages   = {1226--1227},
  year    = {2011},
  doi     = {10.1126/science.1213847},
  url     = {https://doi.org/10.1126/science.1213847}
}

@misc{banbury2021mlperf,
  author       = {Banbury, Colby and Janapa Reddi, Vijay and Torelli, Peter and Holleman, Jeremy and Jeffries, Nat and Kiraly, Csaba and Montino, Pietro and Kanter, David and Ahmed, Sebastian and Pau, Danilo and Thakker, Urmish and Torrini, Antonio and Warden, Peter and Cordaro, Jay and Di Guglielmo, Giuseppe and Duarte, Javier and Gibellini, Stephen and Parekh, Videet and Tran, Honson and Tran, Nhan and Wenxu, Niu and Xuesong, Xu},
  title        = {{MLPerf Tiny Benchmark}},
  year         = {2021},
  eprint       = {2106.07597},
  archivePrefix= {arXiv},
  primaryClass = {cs.LG},
  doi          = {10.48550/arXiv.2106.07597}
}

@article{schwartz2020green,
  author  = {Schwartz, Roy and Dodge, Jesse and Smith, Noah A. and Etzioni, Oren},
  title   = {Green AI},
  journal = {Communications of the ACM},
  volume  = {63},
  number  = {12},
  pages   = {54--63},
  year    = {2020},
  doi     = {10.1145/3381831}
}

@misc{gacrc_sapelo2,
  author       = {{Georgia Advanced Computing Resource Center}},
  title        = {{Systems: Sapelo2}},
  howpublished = {\url{https://wiki.gacrc.uga.edu/wiki/Systems}},
  year         = {2026},
  note         = {University of Georgia}
}

@misc{onnxruntime_tensorrt_ep,
  author       = {{ONNX Runtime}},
  title        = {{TensorRT Execution Provider}},
  howpublished = {\url{https://onnxruntime.ai/docs/execution-providers/TensorRT-ExecutionProvider.html}},
  year         = {2026}
}
